\documentclass[sigconf]{acmart}
\newif\ifdraft\drafttrue

\usepackage{booktabs}
\usepackage{tikz}
\usepackage{amsmath}
\usepackage{amsfonts}
\usepackage{xcolor}
\usepackage{xspace}
\usepackage{enumerate}
\usepackage{verbatim} 
\usepackage{cuted}
\usepackage{subcaption}
\usepackage{soul}
\usepackage[xcolor]{changebar}
\usepackage[scaled]{beramono}
\usepackage{courier}
\usepackage{amsmath,amsthm,mathtools}
\usepackage{url}

\usepackage{listings}
\newcommand{\subsubsubsection}[1]{\smallskip\noindent\textbf{\textbf{#1}.}}

\newcommand{\sys}{ITX}

\setcopyright{none} 
\acmYear{2022}

\begin{document}

\title{Confidential Machine Learning within Graphcore IPUs}

\author{\normalsize Kapil Vaswani,$^1$ Stavros Volos,$^1$ Cédric Fournet,$^1$ Antonio Nino Diaz,$^1$ Ken Gordon,$^1$ Balaji Vembu,$^0$ Sam Webster,$^1$ David Chisnall,$^1$ Saurabh Kulkarni,$^2$ Graham Cunningham,$^2$ Richard Osborne,$^2$ Dan Wilkinson$^2$}
\author{$^1$Microsoft Research $\hspace{0.1in}$ $^2$Graphcore}

\begin{abstract}
We present IPU Trusted Extensions (\sys), a set of experimental hardware 
extensions that enable trusted execution environments in Graphcore's AI accelerators.

\sys\ enables the execution of AI workloads 
with strong confidentiality and integrity guarantees at low performance overheads. 
ITX isolates workloads from untrusted hosts,
and ensures their data and models remain encrypted at all times except within
the IPU.
\sys\ includes a hardware
root-of-trust that provides attestation capabilities and orchestrates trusted
execution, and on-chip programmable cryptographic engines for authenticated
encryption of code and data at PCIe bandwidth. 
We also present software for \sys\ in the form of compiler and runtime
extensions that support multi-party training without requiring
a CPU-based TEE.

Experimental support for \sys\ is included in Graphcore's GC200 IPU taped out at 
TSMC's 7nm technology node.
Its evaluation on a development board
using standard DNN training workloads suggests
that \sys\ adds less than 5\% performance overhead, and delivers
up to 17x better performance compared to CPU-based confidential computing systems relying on AMD SEV-SNP.  
\end{abstract}

\maketitle

\footnotetext{Work done while Balaji Vembu was affiliated with Microsoft.}

\section{Introduction}

Machine learning (ML) is transforming many tasks such as medical
diagnostics, video analytics, and financial forecasting. 
Their progress is largely driven by the
computational capabilities and large memory bandwidth of AI accelerators such as
NVIDIA GPUs, Alibaba's NPU~\cite{alibaba:npu}, Google's TPU~\cite{jouppi:tpu}, and Amazon's Inferentia~\cite{amazon:inferentia}.
Their security and privacy is a serious concern:
%
due to the nature and volume of data required to train sophisticated models, the
sharing of accelerators in public clouds to reduce cost, and the increasing frequency and severity of data breaches,
there is a realization that machine learning systems needs stronger end-to-end
security mechanisms that protect their sensitive models and data.

Confidential computing~\cite{ccc, amazon:cc, google:cc, microsoft:towardccc} relies on
custom hardware support for trusted execution environments (TEE), also known as
enclaves, that can provide such security guarantees.
Abstractly, a TEE is capable of hosting code and
data while protecting them from privileged attackers.
The hardware can also measure this code and data
to issue an \emph{attestation report}, which can be verified by 
any remote party to establish trust in the TEE. 
In principle, confidential computing enables multiple organizations
to collaborate and train models using sensitive data, and
to serve these models with assurance that their data and models remain
protected.
However, existing TEEs such as Intel SGX~\cite{mckeen2013innovative},
AMD SEV-SNP~\cite{amd:sev-snp}, and ARM Trustzone~\cite{trustzone} are
restricted to CPUs and cannot be used for applications that offload
computation to accelerators.

Adding native support for confidential computing into AI accelerators
can greatly increase their security, but also involves many
challenges.
Security features 
such as isolation, attestation, and side-channel resilience
must be fitted in their highly optimized architecture,
with minimal design changes, 
and without degrading their functionality, performance, or usability.
An additional requirement is the flexibility to operate with different 
hosts, including CPUs with no TEE support, CPUs with process-based
TEEs such as Intel SGX, and CPUs with VM-based TEEs such as AMD SEV-SNP. 
Finally, the manufacturing and assembly process and protocols must be hardened against supply chain attacks. 

\begin{figure}[t]
  \center
  \includegraphics[width=\columnwidth]{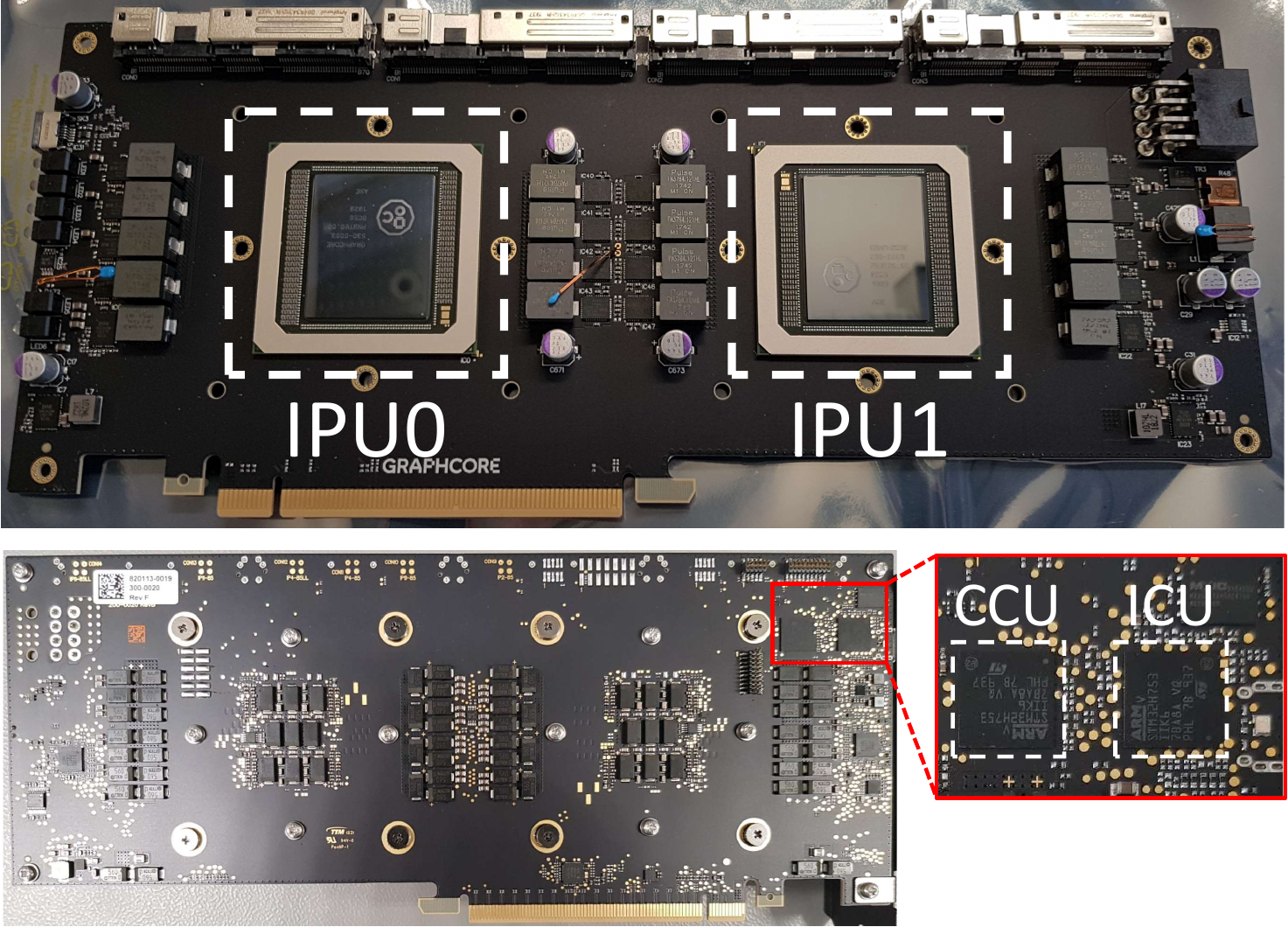}
  \caption{\small GC200 IPU development board with the ITX
    extensions for confidential ML showing the two IPUs on the front
    side connected to the CCU via the ICU (on the back).}
  \vspace{-0.2in}
\end{figure}

This paper describes our effort to support TEEs in a state-of-the-art AI 
accelerator, Graphcore's Intelligence Processing Unit (IPU). 
We introduce \textit{IPU Trusted Extensions} (\sys), a set of experimental
hardware capabilities in the IPU. 
We show that, using \sys\ in conjunction with appropriate compiler and
runtime support, we can delegate ML tasks to the IPU with strong
confidentiality and integrity guarantees while delivering accelerator-grade performance.
In particular, {\sys} can guarantee isolation of an ML application from an untrusted host:
application code and data appears in cleartext only within the IPU, and
remains encrypted otherwise, including when transferred over the PCIe link
between the host and the IPU.
Once an application is deployed within an {\sys} TEE, the
host can no longer tamper with the application state or the IPU
configuration.
{\sys} can also issue remotely verifiable attestations, rooted in
a Graphcore PKI, enabling a relying party to establish trust
in a given ML task before releasing secrets such as data decryption keys. 

The main components of {\sys} are a new  execution mode in the IPU for isolating all security sensitive state from the host and securely
handling security exceptions, programmable cryptographic engines capable
of encrypting and decrypting PCIe traffic between the host CPU and the IPU at
line rate (32 GB/s bidirectional throughput for supporting PCIe Gen4), and a
novel authenticated encryption protocol for ensuring confidentiality and
integrity of code and data transfers without requiring trust in the host. 

Trust in {\sys} is rooted in the \textit{Confidential Compute Unit (CCU)},
a new hardware Root-of-Trust on the Graphcore board.
The CCU provides each device with a unique identity based on a hardware secret
sampled within the CCU at the end of manufacturing.  
The CCU firmware is responsible for managing the entire lifecycle of TEEs on the
IPU, including creation, issuing attestation reports that capture IPU and task
specific attributes, key exchange, launch, and termination of TEEs.
Our design also features protocols for securely provisioning firmware to the IPU
in a potentially hostile manufacturing environment, for issuing certificates
that capture the identity of all updatable firmware, and for supporting
firmware updates without requiring device re-certification.

Several distinguishing aspects of {\sys} and the IPU programming model 
result in stronger security than one may expect from CPU-based TEEs,
notably as regards side channels:
\begin{itemize}
  \item An {\sys} TEE spans the entire IPU, and has exclusive access
  to all IPU resources until it terminates. Therefore, it is not possible for an adversary to
  run concurrently on the same resources and exploit the resulting side
  channels. This execution model is feasible since most AI workloads require at
  least one accelerator, with larger workloads requiring thousands of accelerators for many hours.


\item The IPU's memory system consists of large amounts of 
on-chip SRAM attached to its cores, which is loaded with data from 
untrusted external memory during explicit synchronization phases. Thus, 
during computational phases, code and data accesses to IPU memory 
have a fixed latency.
  This has two security implications:
  (1) traffic between the IPU cores and memory need
  not be encrypted, since it stays within the chip; 
  (2) this avoids the need for optimizations such as caching or
  speculation to hide memory access latency, and the resulting side channels.
  
\item The IPU supports a programming model where allocation and
  scheduling of all resources on the IPU (cores, memory, and
  communication channels) are statically managed by the compiler.
  Hence, the IPU application binary defines its entire data and
  control flow, including data transfers within the
  IPU, and between IPU and host memory.  This is unlike GPUs where the
  host software stack (runtime and driver) remain in full control of
  the execution, and therefore must be trusted to some extent to
  guarantee integrity.
\end{itemize}

There are many ways for software to utilize {\sys} to provide end-to-end guarantees
for ML workloads, depending on the threat model and capabilities of the host. 
This paper focuses on configurations where a multi-party ML
training workload is deployed to the IPU \textit{without trusting the host CPU}.
This mode has the strongest security properties and can be used with any CPU.
We describe a prototype software stack and protocols for it, and present its end-to-end 
evaluation using standard DNN training workloads.
Software to support other configurations, e.g., where the IPU is coupled with
a hardware-protected CPU TEE, are left for future work. 

We have fully implemented {\sys} in the GC200 IPU, manufactured in 
TSMC's 7nm technology.
Our extensions use less than 1\% of this large ASIC, 
and do not require any change to its compute core or memory subsystem.
Its evaluation on a development board using confidential ML training workloads
suggests a performance overhead of less than 5\% compared to non-confidential
IPU workloads.
While our prototype demonstrates promising results, significant work remains to turn our work into production.

Due to implementation constraints, our prototype uses a discrete hardware root-of-trust (instead of an on-die core) and it does not encrypt traffic over IPU-IPU links. 
It is therefore vulnerable to physical attacks, e.g., on the link between the CCU and IPU, or between multiple IPUs. 
These vulnerabilities are not limitations of our design and can be
addressed in future IPU generations by integrating the root-of-trust on the IPU
chip, and introducing additional encryption engines on IPU-IPU links.

In summary, this paper makes the following contributions:

\begin{enumerate}[(1)]
  \item A set of experimental hardware extensions to the IPU, Graphcore's AI
  accelerator, that enable high-performance confidential multi-party machine
  learning. 

  \item Support for remote attestation and secure key exchange based on a
  discrete hardware root-of-trust. 

  \item A pipelined application-level protocol for authenticated encryption \& decryption of code and data over PCIe. 

  \item Protocols for securely provisioning secrets, firmware and certificates to
  a device during manufacturing. 

  \item Prototype software support for enabling confidential multi-party
  training of ML models expressed in TensorFlow on the IPUs without requiring
  trust in the CPU. 

  \item Implementation of \sys\ in the IPU ASIC manufactured by TSMC in 7nm
  technology, and its initial evaluation on a development board, which suggests low overheads
  and orders of magnitude improvements over CPU TEEs. This makes our prototype the first AI accelerator to support confidential computing.
\end{enumerate}

\noindent
While some aspects of our design are specific to Graphcore IPUs, we
hope it can serve as a blueprint for adding TEE support in other
specialized devices and accelerators.

\section{Background}
This section outlines the Graphcore IPU architecture and its programming model, 
with an emphasis on aspects relevant to security. 
A more detailed description of IPUs and a comparison with GPUs are out of scope---see, e.g., 
\cite{graphcore:overview, graphcore:perf}.
The section also reviews hardware-based confidential computing. 


\subsection{IPU Hardware Architecture}
\label{sec:ipu}

\subsubsubsection{Tiles} 
Each IPU consists of a set of \emph{tiles}, each with a multi-threaded core
and a small amount of private on-chip SRAM.
The GC200 IPU features 1472 tiles, totalling roughly 900 MB of on-chip SRAM. 
The cores support an instruction set tuned for AI, including
specialized vector instructions and low-precision arithmetic.
Each core can execute up to six statically scheduled threads.
Since on-chip memory can be accessed at fixed latency, most instructions
can be exactly scheduled by the compiler.
(Other IPU configurations may provide connectivity between tiles and additional on-board DRAM, which would be accessed in a similar way as host DRAM; these configurations are out of scope in this paper.)

\begin{figure}[t]
  \includegraphics[width=\columnwidth]{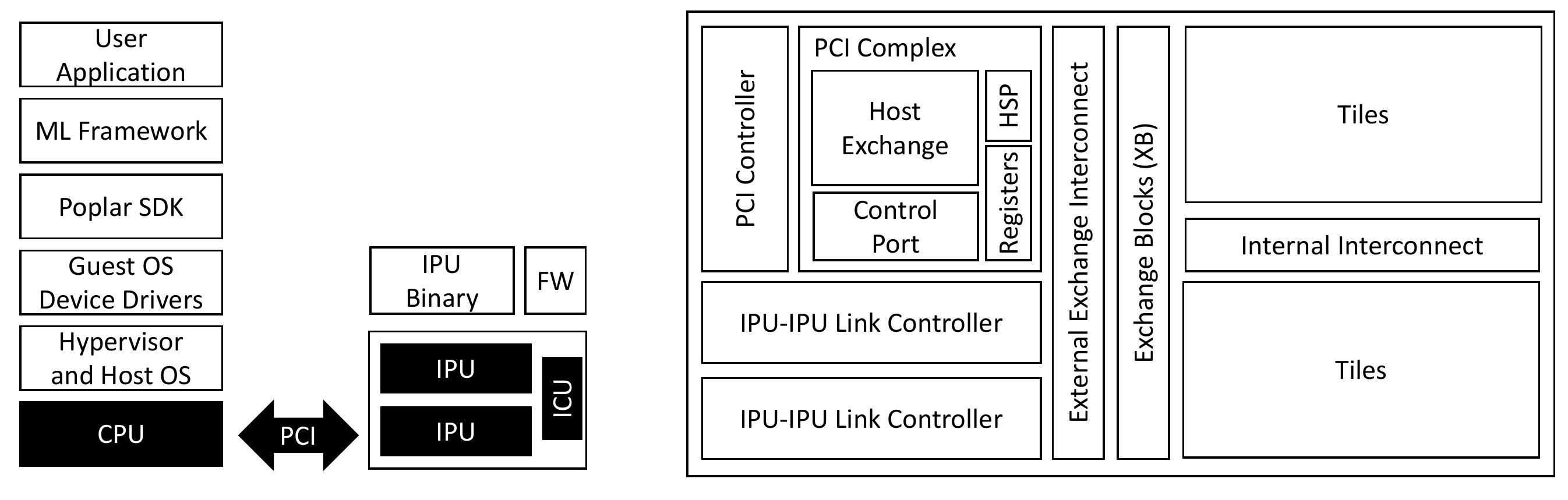}
  \center
  \vspace{-0.15in}
  \caption{\small System stack (left) and IPU floorplan (right).}
  \label{fig:ipu-stack}
  \vspace{-0.15in}
\end{figure}

\subsubsubsection{Interconnects}
The tiles are connected over a high-bandwidth \emph{internal exchange}, an
all-to-all, stateless, synchronous and non-blocking interconnect whose operation
is similarly orchestrated by software.
The internal exchange is connected to an \emph{external exchange} interconnect
via a set of \emph{exchange blocks}. 
Each exchange block manages a subset of the tiles and mediates traffic
between the two interconnects. 
Each IPU has a pair of PCIe links that connect to a host server, and
additional IPU-Links that connect to other IPUs.

The external interconnect is a packet-switched Network-on-Chip. 
Tiles use the external interconnect to dispatch packets to the host via PCIe
links and unicast/multi-cast packets to tiles on other IPUs via IPU-links.
Tiles read data from the host by issuing a \emph{read request} packet and waiting for all associated \emph{read completion} packets. 
Tiles write data to the host by issuing one or more \emph{write request}
packets. 
Packets are routed based on tile identifiers. 
For requests, packets from exchange blocks are placed onto
lanes based on the source tile identifier of the exchange packet.
For read completions, the exchange lane is chosen based on the
destination tile identifier, which is recorded in a lookup table in the PCI
complex for each outstanding read request.

\subsubsubsection{IPU Address Spaces} 
The IPU exposes three address spaces, collectively known as the \emph{IPU
exchange address space}, to facilitate communication between the host and the
IPU and between IPUs. 
The \emph{Tile address space} is used by tiles to address one another.
The \emph{Host PCI space} is used by the host to address tile memory and on-chip
page tables in the Host Exchange block.
The \emph{Tile PCI space} is used by tiles to address read requests to host memory over PCI.
The IPU can be configured to re-map read requests from tiles to the PCI
domain using on-chip page tables. 

\subsubsubsection{Host-IPU Interface}
The IPU exposes a set of configuration registers to the host via a PCI BAR
space.
These registers are hosted in a component known as the \emph{PCI Complex}. 
The PCI complex consists of
a \emph{Host Sync Proxy (HSP)} that is responsible for external synchronization between the host and the IPU, a \emph{host exchange} that translates packets between PCI format and a
proprietary external-exchange packet format,
on-chip page tables for address translation of read/write requests from tiles to the PCI domain,
on-chip lookup tables for keeping metadata for outstanding PCI read requests,
and a \emph{control port} that provides access to configuration registers of all
other internal components.

The host exchange subsystem also includes a component known as the
\emph{autoloader}, which enables efficient scrubbing and initialization of tile
memory. 
To initialize a binary in tile memory, the host can load small programs (e.g., a
bootloader) into the autoloader, which can then broadcast it to
all tiles.

\subsubsubsection{Host-IPU Synchronization} 
The IPU execution model is based on the Bulk Synchronous Parallel (BSP)
paradigm, with barriers and supersteps. 
A superstep involves a global synchronization barrier between all tiles
on one or more IPUs, followed by an exchange phase that transfers
data between tiles, followed by a compute phase which ends at another barrier.
This process repeats until some application specific criteria is met---e.g., loss
is under a threshold. 

Using the Host Sync Proxy registers, the host can configure the frequency of
synchronization barriers and indicate barriers at which it expects to be
notified---e.g., when one or more batch of data has been processed, at epoch
boundaries. 
Once configured, the IPUs can execute multiple supersteps independently without
requiring involvement from the host. 

\subsubsubsection{IPU Control Unit (ICU)}
\label{sec:icu}
The ICU is a microcontroller integrated on the board and connected with the IPUs
via JTAG, and with PCB peripherals for power supply and environmental
monitoring.
It is responsible for initialization and power management of the IPUs. 

\subsubsubsection{Resets} 
The main means of resetting the IPU from the host is a \emph{secondary bus
reset (SBR)} that resets the entire device including the IPUs, the ICU, and the host link; 
the ICU must re-enable the host link once it comes out of reset.
Alternatively, a \emph{Newmanry Reset} can be triggered by writing the IPU control register; it resets the device
logic including the host and IPU links, but does not reset the
physical links.
In both of these resets, tile memory is not scrubbed.

\subsection{IPU Software Stack}
Graphcore provides a software stack, known as Poplar,
for compiling and executing applications written in ML frameworks such as TensorFlow
and PyTorch.
Poplar consists of a compiler, a host runtime, and a set of libraries 
supported by the IPU device driver. 

\subsubsubsection{Compilation} 
The Poplar compiler is responsible for compiling a computation graph representing a task 
(e.g., a TensorFlow XLA graph) into IPU binaries. 
Compilation involves statically partitioning each layer in the computation
graph between tiles, with each tile holding a part of the model state (weights
and activations for some layers) and a part of the input data.
The compiler assigns resources (threads, memory) to each node of the graph, 
schedules its computation, and emits specialized code for each tile. 
%

The resulting IPU binary captures the different phases of execution, including I/O for reading 
batches of data, code for running the training loop, and I/O for writing 
the weights of the trained model. I/O phases also include synchronization 
and internal exchange code for exchanging data among tiles.

The Poplar compiler maps all data transfers between the host and IPUs
to an abstraction called \emph{streams} supported by the runtime. 
Data transfers from the host to an IPU (and IPU to the host) are mapped to input (output) streams
and compiled to sequences of read (write) instructions to the Tile PCI address space. 
%
The compiler also uses streams to implement checkpoints; checkpoint
creation maps all model weights to a single output stream, and 
checkpoint restoration reads them back from a single input stream.

The compiler supports an offline mode, which decouples compilation from execution. 
In this mode, the compiler generates self-contained IPU binaries, which can
be persisted and loaded into one or more IPUs at a later point in time. 

\subsubsubsection{Host Runtime} 
The runtime provides abstractions for loading IPU binaries, and for
streaming data in and out of the IPUs. 

For loading IPU binaries, the runtime deploys a small bootloader into
a reserved section of each tile memory.  The bootloader in turn reads
tile-specific application binary from the host into tile memory.

The runtime implements input streams by repeatedly copying data into
a ring buffer in host memory and mapping the pages of the ring buffer
into Tile PCI space in the on-chip page table.
Once the ring buffer is ready and
the mapping is defined, code on tiles can issue read
requests. Similarly, output streams are implemented by copying data
from the ring buffer to application memory.

\subsection{Confidential Computing}
Confidential computing is a paradigm where code and data remains
protected from priviledged attackers throughout their lifecycle,
including when it is at rest, in transit and \emph{during use}. 
Central to confidential computing is the notion of a \emph{trusted execution
environment} (TEE). 
TEEs offer two key capabilities: the capability to
host an application in a hardware-isolated secure environment, which protects the
application from all external access including access from privileged
attackers; and the capability to issue remotely verifiable attestations, which
capture various security claims about the application hosted in the TEE and the
platform supporting the TEE. 
These attestations can be used by any relying party to establish trust in an
application and opening secure channels for communication. 

TEEs are supported by recent processors from Intel and AMD. ARM
has recently defined a specification for supporting TEEs.
There are broadly two classes of CPU TEEs: process-based and VM-based. 
Process-based TEEs (e.g., Intel SGX) are designed to isolate a user-space
application from an untrusted operating system (both guest and host) and the hypervisor. 
VM-based TEEs (e.g., AMD SEV-SNP, Intel TDX) are designed to protect an
entire guest VM from the host operating system and the hypervisor. 
TEEs offer varying degrees of protection from attackers with physical
access to the CPU. 
Most TEE implementations assume that attackers can snoop on
interconnects between the CPU package and external components (e.g., off-chip
DRAM) and protect data by encrypting and integrity-protecting memory traffic. 
Information leakage through side-channels is still often considered out of scope,
although CPU vendors are offering defense-in-depth protection against
specific side-channels, such as those based on speculation.

Support for remote attestation is typically rooted in an on-die hardware root of
trust (HRoT), which has exclusive access to a unique device secret
provisioned into one-time programmable fuses during manufacturing.
During boot, the HRoT uses the secret to derive a device-specific identity key.
The corresponding public key is endorsed by the hardware manufacturer. 
This key typically endorses keys used for signing attestation reports
for a TEE. 

\section{Threat Model}\label{sec:threat}
TEE hardware is subject to a variety of attacks throughout its lifecycle, from
chip design and manufacturing up until the hardware is decommissioned.

Trust in TEEs is rooted in hardware, and consequently in the chip designers and their
OEMs involved in designing and manufacturing the chips. 
Additional trust is also required in the infrastructure for issuing certificates
to each chip, and for publishing the last known good version of firmware TCB.
While this is also the case with the IPU, we wish to minimize trust in the rest
of the supply chain.
Hence, we conservatively assume that attackers control the manufacturing and assembly
process after tapeout, including the process of provisioning firmware and/or
secrets to each device and harvesting their Certificate Signing Requests (CSR).

After deployment, we assume a strong adversary that controls the entire system
software stack, including the hypervisor and the host operating system, and also has
physical access to the host. 
The adversary can access or tamper with any code and data transferred between
the host CPU and the IPU, either in operating system buffers or over PCIe. 
The adversary can also tamper with device memory directly via the PCI BAR,
or map the victim application's tile PCI address space to host-side memory
controlled by the attacker.
Information leakage through side channels such as traffic analysis, power
consumption, timing, and physical probes on the IPU are generally out of scope. 
However, we do wish to offer protection from side channels
based on memory access patterns, and from low level integrity attacks such as glitching. 

We trust the IPU and the HRoT packages, and we assume that the adversary
cannot extract secrets or corrupt state within the packages.
In particular, the IPU package includes trusted SRAM within the IPU tiles
accessed only via on-chip private channels.

The ML source script and high-level configuration are trusted.
The ML framework and the Poplar compiler are trusted for integrity of the
computation---i.e., to compile the model defined in the ML script correctly
into a manifest and binaries that run on the IPU.

In multi-party configurations (involving parties that do not trust one
another), these assumptions can be addressed by having all parties
review the script and configuration for the workload, then confirm
that they all locally compile to the same manifest and binaries.
Each party is trusted with the integrity and confidentiality of the
data streams they provide for the computation; in particular, honest
parties are trusted to correctly encrypt their data streams with a
fresh data encryption key, and to release this key to IPUs only after
verifying their attestation report.

In configurations that couple the IPU with a host CPU TEE (e.g., Intel SGX and TDX, AMD SEV-SNP), the CPU package is
also trusted, along with any software hosted in the TEE; we omit the
details of their platform-specific threat models.
With process-based TEEs, such as Intel SGX,
the CPU-based software TCB may include the ML training or inferencing script, ML framework (e.g., TensorFlow, PyTorch), the Poplar compiler and runtime.
With VM-based TEEs, such as AMD SEV-SNP, the TCB may additionally include the Poplar 
kernel-mode driver and a guest operating system.
The Poplar runtime is then trusted for confidentiality---i.e., to
setup a secure, attested channel between the CPU TEE and IPU,
and to transfer code/data over this secure channel.

With the current generation of IPUs, we make additional trust assumptions in the ICU, which
provides connectivity between the hardware RoT and the IPUs, and in links
between IPUs. 
We trust the ICU firmware and the physical links that connect the HRoT, the ICU
and the IPU. 
These trust assumptions can be removed in subsequent generations of the IPU
by placing the HRoT on the IPU die, and encrypting communication over IPU-IPU links. 

Under this threat model, we wish to provide confidentiality and integrity guarantees for 
model code and data, including initial weights, input data, checkpoints and outputs. 
For training, integrity implies that the final outcome (i.e., the
trained model) is bitwise equivalent to the model obtained in the absence of the
attacker. 
For inferencing, integrity implies that requests yield the same results as
those obtained in the absence of the attacker.
Conversely, liveness properties (e.g., progress or availability) are out of scope.

We wish to also
provide remote attestation, which refers to the ability of the platform to make remotely
verifiable claims that a relying party can use to reason about the TEE's security
properties and thereby establish trust in the application hosted
within the TEE even in the presence of an attacker.
Specifically, we wish to ensure that the attestation can deliver temporally
fresh evidence that contains all security-sensitive parts of the platform
and application state, and that the underlying attestation mechanism is 
trustworthy and robust to advanced attacks such as chosen-firmware attacks. 

\section{Overview}
\label{sec:overview}

Trusted execution in IPUs enables model developers
to securely offload an ML job (training or inferencing)
while protecting both model and data from the hosting platform. 
In turn, model developers can prove to data providers that their data remains
protected from both the hosting platform and the model developers themselves. 

The workflow for securely offloading an ML job involves multiple
steps, starting with the creation of a TEE, generation of an
attestation report, its verification by remote parties (e.g., the model
developer or data providers), encryption of code and data, secure exchange of
encryption keys with the IPU, job execution, and decryption of
the outcome: a trained model or inference results.

\subsection{Hardware extensions (\sys)}
The IPU hardware contains several components (shown in
Figure~\ref{fig:ccu-sxp-integration}) to support this workflow. 

First,  a new hardware root-of-trust integrated on
  the IPU board, called the \emph{Confidential Computing Unit (CCU)}.  
  The CCU gives each board a unique identity based on a 
  hardware secret generated by the CCU during manufacturing.
  The CCU firmware supports an API which an untrusted host can use to manage
  the entire TEE lifecycle on the IPU, including creation, attestation, key
  exchange, application launching, and termination.
  Section~\ref{sec:rot} describes the architecture of the HRoT and its role in trusted execution.

Second, a new mode, called the \textit{trusted mode}, in which all
  security sensitive state is isolated from a potentially malicious host.
  This mode is entered by writing to a configuration register.
  (For remote verifiability, this register is measured by the CCU and
  included in the attestation report.)
  Once the IPU enters this mode, its configuration registers and tile memory can
  be accessed only by the CCU and ICU.
  The only way to exit this mode is via a chip reset,
  which is extended to scrub all key registers and tile memory.
  
Third, programmable AES-GCM engines for authenticated encryption and
  decryption of code and data transferred between the host and the IPU at PCIe
  line rate.
  These engines are hosted in new components, called \emph{Secure
  Exchange Pipes (SXP)}, located on the interconnect between the PCIe
  block and the exchange blocks. 
  The SXP and its use are described in Section~\ref{sec:auth-enc}. 

\begin{figure}[t]
  \includegraphics[width=\columnwidth]{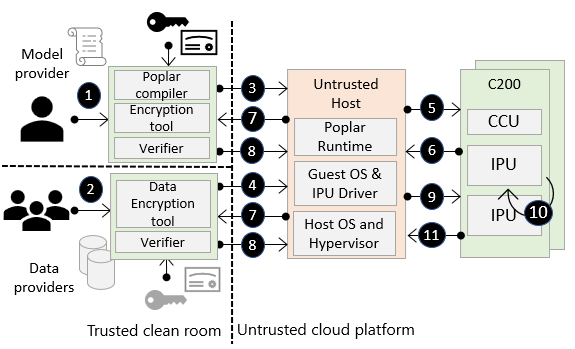}
  \vspace{-0.25in}
  \caption{ \small
    Multi-party training in trusted offline mode.
    Before training, the remote parties upload their encrypted code, data and
    certificates (1--4)
    Once training starts (5--6), they verify the attestation report (7)
    then release their encryption keys to the CCU (8-9); 
    they can be offline for the rest of the computation.
    The IPUs train the model in a TEE (10) and releases an encrypted trained
    model (11), whose key can be shared with model receiver(s).}
  \label{fig:multi-party-flow}  
  \vspace{-0.15in}
\end{figure}

\subsection{Software Support}
There are many ways for software to utilize \sys. 
For this paper, we illustrate a particular mode, which we refer to as the
\emph{offline} mode (Figure~\ref{fig:multi-party-flow}). 
In this mode, a multiparty ML training workload can be deployed in an IPU-based TEE \emph{without requiring a CPU-based TEE}.
This mode has strong security properties (e.g., small TCB) and minimal dependencies on the host server hardware.
We discuss limitations of this mode, and extensions for scenarios
such as aggregation and pre-processing, and inferencing in Section~\ref{sec:sw-disc}. 

\subsubsubsection{Job Preparation}
In offline mode, a model developer uses an extended Poplar compiler to
statically compile a model training job expressed in an ML framework such
as TensorFlow or PyTorch to standalone IPU binaries in a trusted, offline
\emph{clean room environment} (1).
In addition to the binary, the compiler generates a \emph{job manifest}, which
contains auxiliary information required at runtime to execute the job. 
Next, the model developer encrypts binaries and parameters such as initial
weights and learning rate using encryption keys that remain in the clean room
environment. 
The model developer also generates a fresh public key share for key exchange,
and a signature over the key share using their certificate. 
These artifacts, along with the model developer's certificate are packaged together to create an \emph{application package}.
Separately, data providers pre-process and encrypt their input data and labels
in their own clean room environments, and create \emph{data packages} which
include their key shares and certificates (2).
The resulting packages are uploaded to a server with IPUs attached (3, 4).

\subsubsubsection{Job Initialization}
Any entity (including the model developer) can
initiate execution of the training job using the Poplar runtime, which we extend
to load encrypted code and data into the IPUs. 
For confidential computing jobs, the Poplar runtime provides 
user-mode APIs for operations such as creating TEEs (5) for a job, requesting
for attestation reports and additional collateral such as device-specific
certificates (6), and relaying key-exchange messages from relying parties to the CCU (8).
This runtime is not trusted. 

\subsubsubsection{Remote Attestation}
In trusted mode, the CCU can issue remotely verifiable attestations, which are
relayed to relying parties (7) as proof of TEE configuration for their workload.
The attestation is a certificate chain from the Graphcore root CA to an
end-certificate signed by the CCU with custom extensions that embed
initialization attributes (e.g., measurement of all security-sensitive IPU
registers) and job-specific attributes, such as the measurement of the job
manifest, and the hash digest of other runtime attributes, including certificate 
fingerprints of all parties and the CCU's fresh public keyshare. 
The model developer and data providers verify this report, the model, and
identities of other participants.
If they decide to make their data available for this job, they derive shared
encryption keys using the CCU's public key share and securely exchange their
secrets with the CCU (see Appendix~\ref{appendix:job}).

\subsubsubsection{Job Execution}
After the model developer and data providers have relayed their keys to the CCU,
the CCU deploys the keys into the SXPs and starts the job (9) by installing a
bootloader into the IPU tiles using the autoloader. 
The bootloader is designed to fetch the application binary from host memory to
each tile in 1KB blocks. 
In trusted mode, the blocks are decrypted and integrity checked by the SXPs
before being written to tile memory (see Section~\ref{sec:bootstrapping}).
Once the application binary has been transferred, the Poplar runtime initiates execution of the job. 
During execution, tiles generate read requests for data, also in blocks of 1KB.
In trusted mode, the blocks are fetched from host memory over PCIe, and
decrypted and integrity checked by the SXPs before being written to tile memory.
Similarly, all write requests (e.g., checkpoints and trained model) are
encrypted and extended with authentication tags before being written to host
memory. 
The encryption/decryption protocol is mostly transparent to the compiler, which
can compile \emph{any training algorithm} into binary relying on the data being
in tile memory in cleartext and utilizing all compute resources available on the
IPU.
Finally, the IPU encrypts the trained model with a key made available only
to the model receivers listed, such as one or more of the parties involved, or another CPU/IPU TEE, e.g., for inference.

\section{Trusted Execution on IPUs}
\label{sec:trusted}


\subsection{Confidential Compute Unit (CCU)} 
\label{sec:rot}
The CCU is responsible for associating
each Graphcore device with a unique cryptographic identity and
managing trusted execution in its IPUs. The CCU is a discrete chip based on STMicro's STM32H753 microcontroller~\cite{stm32h573}.
This chip was selected as the root of trust based on several security features
required to implement measured boot and offer protection from a variety of attacks throughout the IPU
lifecycle,
such as the abilities to provision a custom bootloader during
manufacturing in a region of one-time programmable flash memory,
and to switch the microcontroller into a mode that prevents external access via interfaces such as JTAG. 

\begin{figure}
  \center 
  \includegraphics[width=.8\columnwidth]{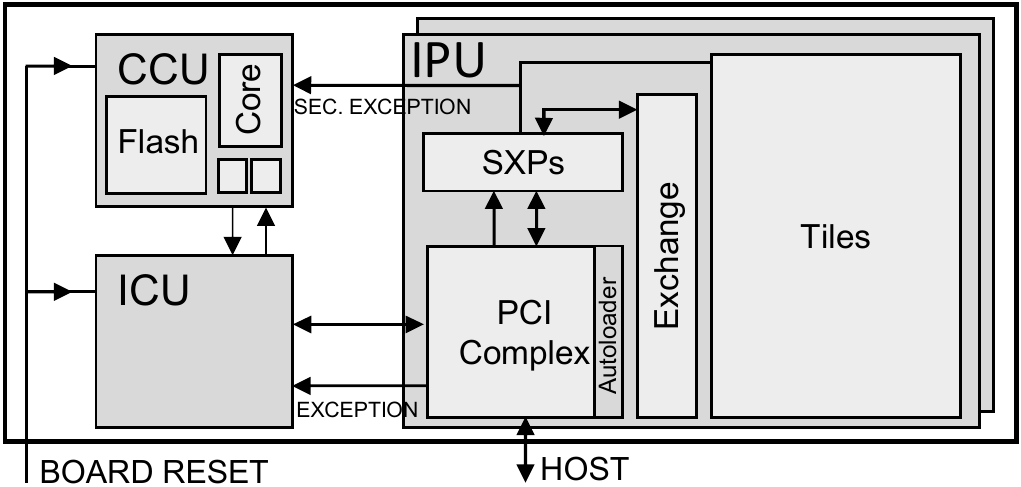}
  \vspace{-0.05in}
  \caption{\small IPU hardware extensions to support trusted execution.}
  \label{fig:ccu-sxp-integration}
  \vspace{-0.15in}
\end{figure}

As shown in Figure~\ref{fig:ccu-sxp-integration}, the CCU is connected to
the IPU via the ICU.
A dedicated pin receives all exceptions generated by the
IPU in trusted mode,
giving the CCU firmware full control over exception handling.
The CCU reset pin is coupled in hardware with the ICU reset pin and  IPU
reset, so they cannot be independently reset.

\subsubsubsection{Firmware Architecture and Attestation}
The CCU implements a measured boot protocol based on the Device Identity
Composition Engine (DICE) architecture~\cite{dice-tcg, dice-engine-spec}.
The protocol is designed to ensure that each device is assigned a unique
identity while minimizing exposure of hardware secrets. 
The protocol also ensures that, except for the stable device identity,
all derived secrets and keys automatically change when firmware (and its
measurement) changes, which ensures that low level firmware attacks such as
boot-kits do not compromise secrets used with other firmware.

\begin{figure}
  \centering
  \includegraphics[height=3in]{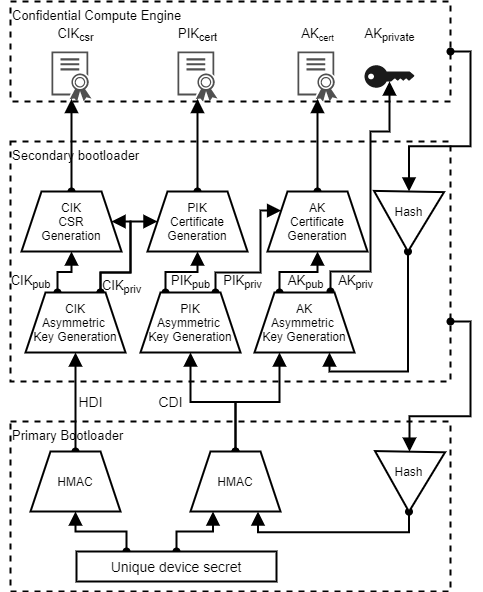}
  \caption{\small CCU firmware architecture and key hierarchy}
  \label{fig:fw-keys}
  \vspace{-0.15in}
\end{figure}

The CCU firmware (Figure~\ref{fig:fw-keys}) consists of three layers:
  an immutable primary bootloader provisioned in one-time programmable
  flash memory at manufacturing;
  a mutable secondary bootloader responsible for device
  identity and attestation certificates; and 
  a confidential compute engine (CCE) that manages the TEE lifecycle.

During manufacturing, the CCU would be provisioned with the primary 
bootloader firmware.
When the device is brought out of reset for the first time, 
this primary bootloader receives control from ROM firmware, samples a
\textit{unique device secret (UDS)} using a hardware-based TRNG,
stores it in a region of flash memory, that is permanently 
and permanently blocks access from any other firmware layers.
The UDS is the root of the IPU's key hierarchy, and this protocol ensures that
it is never exposed outside the CCU, not even to the manufacturer. 

On every subsequent boot, the CCU follows a variant of the DICE measured boot
protocol~\cite{dice-impl}\cite{riot-cert}:
the primary bootloader receives control out of reset, 
loads and authenticates the secondary bootloader from flash using Graphcore firmware signing
key deployed within the primary bootloader.
Next, it derives two intermediate secrets: 
a \textit{Hardware Device Identifier} (HDI) from UDS, and
a \textit{Composite Device Identifier} (CDI) from UDS and the measurement of the secondary bootloader. 
The HDI is unique to each card, whereas the CDI is unique to each
card and secondary bootloader.
It then scrubs any copies of UDS from memory, and transfers control to the
secondary bootloader, handing over both HDI and CDI.

The secondary bootloader further derives two public-private key
pairs:
a \textit{Card Identity Key} (CIK) from HDI, and
a \textit{Platform Identity Key} (PIK) from CDI.
The CIK gives each card a stable identity whereas the PIK is unique to each
card and secondary bootloader.
The bootloader also generates a self-signed CSR for the CIK, a PIK CSR, and a
PIK certificate signed by CIK. 
The PIK CSR and certificate contain a custom extension that records measurements of the secondary bootloader and the ICU firmware along with
additional device-specific information. 
The CSRs would be securely harvested during manufacturing and processed by a Graphcore CA, which would issue CIK and PIK certificates. 

\label{sec:attestation}
The secondary bootloader derives another key pair:
the 
\textit{Attestation Key} (AK) from CDI and the CEE measurement.
Hence, AK is unique to each device, secondary bootloader and CCE.
The bootloader issues an AK certificate signed by PIK; this certificate logs the CCE 
measurement in a custom extension.  
Finally, the bootloader scrubs all secrets and transfers control to CCE,
handing over AK. 

The CCE uses AK to sign attestation reports that contains IPU-specific
information and job-specific information (Section~\ref{sec:tee}). 
A relying party can validate attestation reports using the AK certificate
issued by the device, and CIK and PIK certificates that would be issued by Graphcore. Graphcore would also issue certificates containing the measurements of 
the latest known good CCU firmware and IPU configuration, which a relying party can use
to verify contents of the PIK and AK certificates. 

\subsubsubsection{Firmware Update}
As intended with DICE, a secondary bootloader update invalidates PIK certificates issued by the
manufacturer and, as UDS is provisioned within each device,
it is not possible for Graphcore to independently derive and certify the updated PIK.
Instead, we rely on CIK, acting as a local CA, to sign the updated PIK certificate.
Additionally, Graphcore would issue TCB update certificates containing measurements 
of old and new versions of firmware. 
A relying party can validate attestation reports using PIK certificates
obtained from the device, the original PIK and CIK certificates, and 
TCB update certificates (see Appendix~\ref{app:fw-sign-updates}.)

Note that the protocol above is still susceptible to advanced
chosen-firmware attacks: a malicious secondary bootloader could
impersonate another version of the firmware by using CIK to endorse a
PIK certificate for the corresponding firmware measurement.
Firmware authorization provides a strong defense against such
attacks---the malicious firmware would need to be correctly signed by
Graphcore to run as secondary bootloader. 
We can harden the protocol further by moving CIK and PIK generation
into the primary bootloader, at the cost of increased complexity in one-time 
programmable firmware. Appendix~\ref{sec:hardened-protocol} discusses this variant.

\subsection{TEE Management}
\label{sec:tee}
The CCU exposes an API for creating and managing TEEs on the IPU, outlined below.

\subsubsubsection{TEE Initialization} 
The first step in securely offloading a job to an IPU is to
create a fresh TEE for this job. 
TEE initialization requires a job manifest, public key shares, signatures over
the key shares and certificates for each relying party and a checkpoint counter
indicating whether job is starting or resuming  from a checkpoint. 
During TEE initialization, the CCU first \textit{quiesces} the IPU, ensuring
that there are no in-flight read and write requests between the host and IPU. 
It then switches the IPU into trusted mode, scrubs all tile memory using the
autoloader, and measures the state of the configuration registers. 
It then checks the signatures over the key shares using the certificates.  and
generates its own fresh EC share, which is used to establish  a ECDH shared
secret between each relying party and the CCU (device).

The CCU generates an attestation report signed by the attestation key
containing various IPU-specific attributes, such as configuration register 
measurements, and job-specific attributes, such as the job manifest, certificate
fingerprints for all parties, and the epoch counter and checkpoint identifier 
(see Appendix~\ref{sec:attestation-details} for the details.)

Each relying party (model or data provider) can review the attestation
report, along with the supporting certificate chains, to validate the
device and the initial state of the CCU and IPU, then it can compute
the ECDH shared secret to wrap a key 
package that contains the party's encryption keys for data 
it contributes to the job and nonces (see Table~\ref{tab:crypto} and Secure Key Exchange in 
Appendix~\ref{appendix:attestation-sk} for the details.) 

\subsubsubsection{TEE Launch} After gathering wrapped encryption keys
from all relying parties, the host launches the execution of a job,
which proceeds in several steps.

First, the CCU computes the ECDH shared secret for each party 
and uses them to unwrap the key package(s) received from each party.
It then combines the nonces to derive a checkpoint key and a
final-model encryption key for this run of the job (and, if resuming from another
run, the checkpoint key from that previous run to restore its state). 
This key derivation ensures both that the checkpoint key for this run is fresh (as long as one relying party's nonce is fresh)
and that the checkpoint key of a prior run can be recomputed once all relying parties agree to resume from a checkpoint. Table~\ref{tab:crypto} and Appendix~\ref{appendix:attestation-sk} provide details about the key derivations.
%

Next, the CCU deploys a pre-defined bootloader on the IPU tiles using the
autoloader, and it deploys a first set of encryption keys to the SXP (including the code encryption key)
as specified in the job manifest. 
It then activates the bootloader (whose measurement is included in the 
attestation report) on every tile,
which issues requests to read their encrypted application binary from host memory.
Responses to these read requests are authenticated and decrypted by
the SXPs before being copied into private tile memory.

Finally, The CCU deploys the next set of encryption keys (including
data keys, and possibly the checkpoint key for resumption), as
specified in the job manifest,
and trigger the main execution loop on the IPU tiles.
The CCU may be similarly involved at some synchronization points later
in the job, to deploy different sets of encryption keys.

\subsubsubsection{TEE Termination} 
At any point after initialization of a TEE, the host runtime can also request
that the TEE be terminated. 
The CCU may also trigger TEE termination in the event of a security exception
raised from the IPU e.g., failure to authenticate a response of a read request. 
During TEE termination, the CCU quiesces the IPUs and scrubs tile memory using
the autoloader and disables all keys in the SXPs. 
Finally, the CCU issues a \textit{Newmanry} reset, which switches the IPU back into normal
mode. 

A TEE may be abnormally terminated due to a hard reset of the device.
In such a scenario, the IPU reverts to normal mode and all CCU state is cleared.
When the IPU comes out of reset, prior to re-enabling the host links, the ICU is programmed 
to scrub tile memory to ensure that any secrets left over from a previous execution are erased 
before the host re-gains access to the device.

\section{Encrypted Direct Memory Access}
\label{sec:integrity}
Next, we describe the \sys\ protocol for encrypted code and data transfers to
and from IPU tiles. 
The protocol is designed to ensure confidentiality and integrity even in the
presence of privileged attackers that control the host software stack and can
observe/tamper with PCIe traffic.
The protocol is application-level as opposed to transport-level
(discussed in Section~\ref{sec:related}). 
While it is transparent to ML frameworks, it relies on application software (e.g., the Poplar compiler) to assign IVs for authenticated encryption and for
programming the tiles to securely load code, initial weights, training data, and save/reload checkpoints and results. 
The protocol is supported in IPU hardware by fully pipelined AES256-GCM
engines for authenticated encryption at PCIe line rate. 
This design choice results in simpler hardware (at the cost of some software
complexity), allows the IPU to be coupled with untrusted CPUs or CPUs with
varying TEE support, and retains the compiler's ability to maximize PCIe utilization 
by parallelizing data transfers across multiple tiles. 

\begin{figure}[t]
    \center
    \includegraphics[width=2.75in]{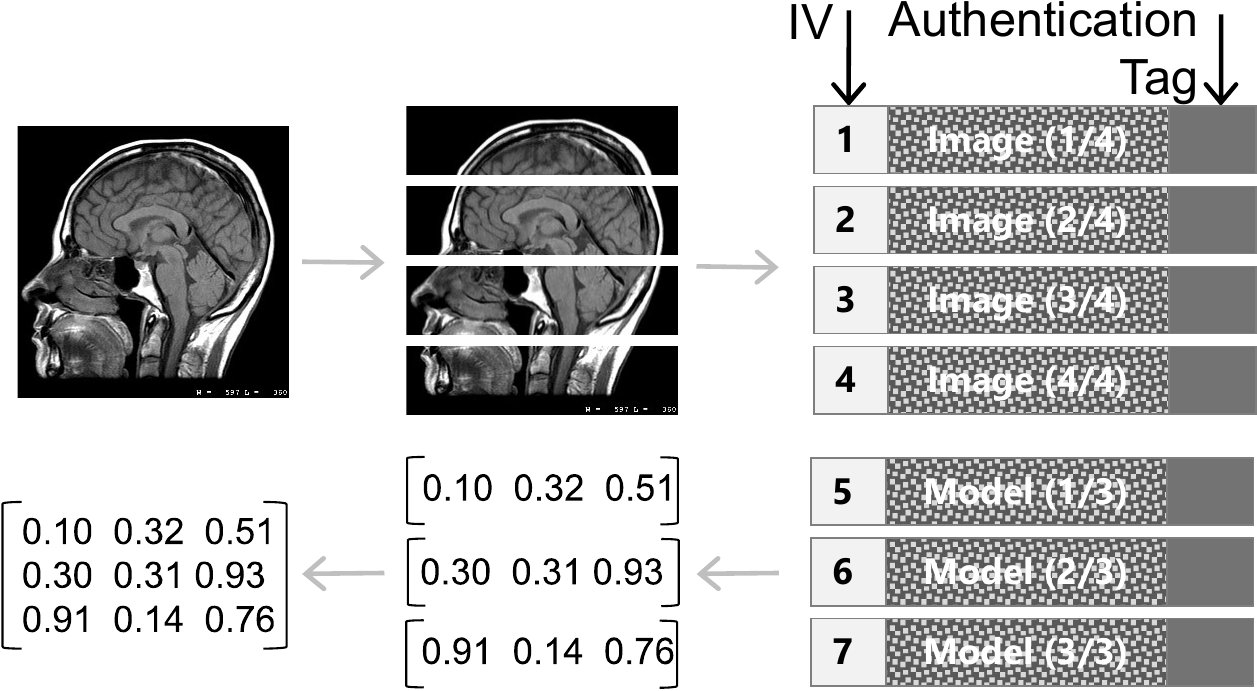}
    \vspace{-0.05in}
    \caption{\small
      Authenticated encryption with explicit IVs.
      Data is statically partitioned into frames with unique IVs.
      Hardware decryption ensures their integrity based on their
      authentication tag; the receiver must verify that received IVs 
      match the expected IVs before accepting the decrypted payloads.}
    \label{fig:sw-encryption-protocol}
    \vspace{-0.15in}
\end{figure}

\subsection{Data Format}
\label{sec:auth-enc}
In the encryption protocol (illustrated in
Figure~\ref{fig:sw-encryption-protocol}), application software partitions each
code and data stream into equally-sized encrypted \emph{frames}.
Each frame consists of a 128-bit IV, followed by a series of cipher blocks that
carry the encrypted contents of the frame, and by a 128-bit authentication tag.
Application software is free to use different frame sizes for different streams,
as long as the total frame size (including IV and authentication tag)
is a multiple of 128 bytes with a maximum of the 1kB, which is the largest
supported PCIe read. 
Application software can use different keys to encrypt different streams. 
This is critical for multi-party scenarios where streams are provided and
accessed by different parties. 
Crucially, application software must ensure that IVs are never reused across
frames encrypted with the same key (which would be catastrophic with AES-GCM). 
In our implementation, this invariant is ensured by the compiler, which
constructs the IV by combining stream-specific identifiers and frame indexes,
and the fact that in Poplar, both code and data
streams are write-once abstractions. 
Together, this guarantees that unless the associated key has been compromised,
authenticated decryption with the correct IV yields the correct payload. 

\subsection{Hardware Support}
Multiple components in the IPU support \sys\ encryption.
The IPU includes blocks, called \emph{Secure Exchange Pipes},
extensions to packet formats for carrying encryption-related information, and
extensions to exchange blocks and the PCIe complex for supporting the task of
mapping frames to keys.

\begin{figure}[t]
    \center
    \includegraphics[width=2.5in]{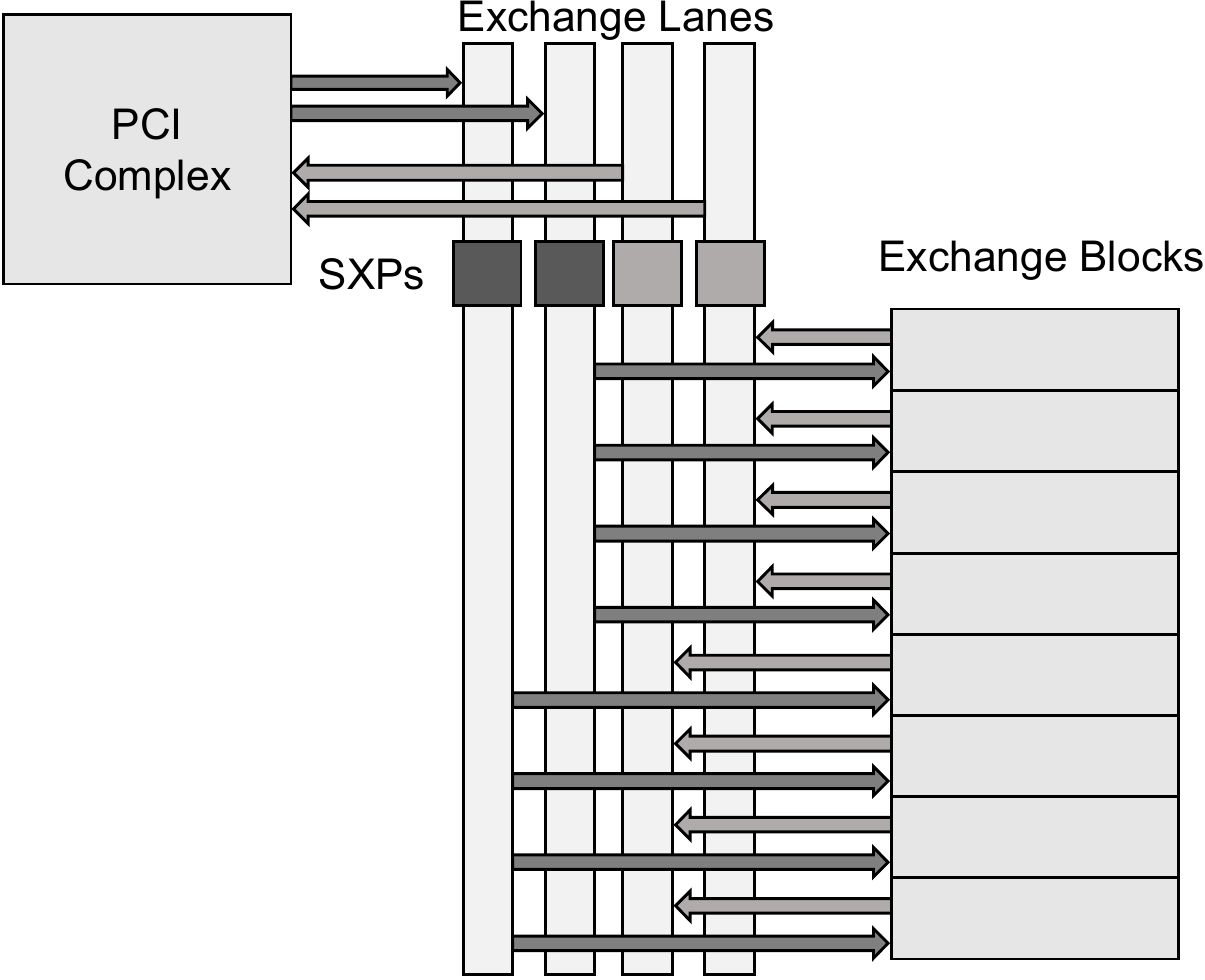}
    \vspace{-0.05in}
    \caption{\small IPU External Exchange Interconnect, with an SXP on each exchange lane.
      Traffic is forwarded from and to exchange blocks to exchange lanes based on the exchange block identifier.}
    \label{fig:sxp-exchange}
    \vspace{-0.15in}
\end{figure}

\subsubsubsection{Secure Exchange Pipe (SXP)}
The SXP is a programmable hardware block that supports AES256-GCM authenticated
encryption and decryption of frames. 
Each SXP achieves 16 GBps unidirectional throughput with
negligible impact on latency. 
As shown in Figure~\ref{fig:sxp-exchange}, there are four SXPs placed on
the exchange interconnect (two per direction) to support
encryption/decryption at PCIe Gen4 line rate (32GBps bidirectional). 
In trusted mode, each SXP is configured to intercept read/write 
requests from four exchange blocks.

\subsubsubsection{AES-GCM Engine} 
The SXP's core is a fully pipelined AES-GCM engine that supports 16 \emph{physical key 
contexts} to enable concurrent requests. 
Each context can be programmed by loading a 256-bit key
into control registers exposed to the CCU via an internal control bus. 
While frames may be interleaved, for functional correctness we require
that \emph{each context processes a single frame at a time}.
This invariant is enforced by the compiler, as detailed in Section~\ref{sec:conf-streams}.

The core implements the standardized AES256-GCM algorithm with two restrictions:
the additional authenticated data is always empty; and the plaintext is
block-aligned and not empty.  
For convenience, we also treat the IV as a full 16-byte block, including the 32
bits of internal block counter.
For each context, the SXP stores
an AES key,
the Galois-hash authentication key (AK) computed by encrypting $0^{128}$ when loading the AES key,
the encryption of the initial IV (EK),
the current IV including the block counter,
and a partial Galois hash (H). 
%
%
%
In each cycle, the core performs one of the following operations on its context:
\begin{itemize}
\item 
The context is idle (i.e., no frame is being encrypted or decrypted) and the core receives the IV for the frame.
It  combines IV with the initial block counter (0), encrypts it and stores the
result in EK; outputs IV; increments IV; and marks the context as active.

\item 
The context is active and the core receives a block of data
. It encrypts IV and XORs it with the data; accumulates this cipherblock
into the partial hash H using AK; outputs the cipherblock; and increments IV. 

\item 
The context is active and the core receives a MAC. It computes the
authentication tag for the frame using AK, EK, IV, and H; compares it with the
received MAC (if performing a decryption); outputs the authentication tag; and
marks the context as idle.
\end{itemize}

The core detects context switches by comparing key context identifiers between
consecutive cycles (as discussed below), so that it can fetch the next context before the next operation.

\subsubsubsection{Frame encryption/decryption}
The SXP receives three types of external exchange packets: read requests
(egress), read completions requiring decryption (ingress), write requests requiring encryption (egress).
Their headers are extended to carry additional
information to help the SXP determine how the packets should be handled: 
  an \texttt{AES} bit indicates that the read completion or the write request is encrypted;
  a 4-bit \texttt{KEY\_INDEX} field identifies the physical key context to use; and
  a \texttt{CC} bit indicates the last packet of the frame and
  triggers the computation of its authentication tag.

In write request packets (outbound to the host PCIe domain), the \texttt{AES}
bit and the \texttt{CC} bit are set by the tile, whereas the \texttt{KEY\_INDEX}
is set by the SXP (as described below). 
In read completions packets, the information is set by the PCIe complex based on trusted state it maintains about pending read requests (as described below). 

Read request packets and packets with the \texttt{AES} bit unset do
not require encryption/decryption; they are passed unchanged.
For all other packets, the header bypasses the AES core, then 
\begin{itemize}
\item 
If the packet is the start of a new frame, the first block (16
bytes) of payload is passed to the AES core as $\texttt{AES\_IV}$, and
the subsequent blocks are passed as $\texttt{AES\_DATA}$.
  
\item 
If the packet is the continuation of a frame, its blocks of payload are passed as $\texttt{AES\_DATA}$.
  
\item
If the packet is the end of a frame (flagged by the header \texttt{CC} bit),
all but the last block of payload are passed as $\texttt{AES\_DATA}$, and the
last block is passed as $\texttt{AES\_MAC}$. (This block carries the
authentication tag when decrypting, and a padding block otherwise.)
\end{itemize}
\noindent 
Packets that both start and end a frame are similarly handled. 

\subsubsubsection{Key Selection}
As described earlier, each SXP supports multiple physical key contexts to enable
encryption/decryption of concurrent I/Os. 
The SXP provides a set of registers that can be programmed by the CCU
to define a mapping between packets and the physical key
context to use for encrypting/decrypting their payload. 
The registers, known as the \emph{exchange block context map} (\texttt{KXBCTXMAP}), define a mapping from exchange block contexts to physical
key contexts.
The compiler can define this mapping by assigning a set of tiles associated
with an exchange block context to read/write from/to a single stream. 
When the SXP receives a packet, it computes the index of the exchange block
context from the source tile identifier in the header, and uses it to look up
the physical key context in \texttt{KXBCTXMAP}. 

As additional defense-in-depth against misconfiguration or corruption of these
registers, the SXP defines two additional sets of registers. 
The first set of registers, known as the \textit{key region definition}
registers (\texttt{KSELLIMIT}), 
can be used to define up to 17 disjoint tile PCI address regions, with the
expectation that  the compiler will allocate streams encrypted with different
keys in different key regions. 
The first region 
is always interpreted as the cleartext region, and is used to map public data;
reads from or writes to this region bypass the SXPs.
A second set of registers, known as \textit{physical context map}
(\texttt{KPHYSMAP}) registers, define a 1:1 mapping between physical key
contexts and key region identifiers.
After inferring the physical key context using \texttt{KXBCTXMAP}, the SXP
looks up \texttt{KPHYSMAP} using the physical key context to obtain a
logical key region identifier.
Then, it looks up the key region definition registers to obtain the key region.
Finally, it checks if the tile PCI address of the request belongs to the region. 
A failure of this check indicates a misconfiguration of either one of these sets
of registers, and causes the SXP to generate a security exception. 

Once the SXP infers the physical key context, it updates
the \texttt{KEY\_INDEX} field in the request packet header. For write 
requests, the field is then used by the SXP to switch the AES core to 
the inferred physical context for encrypting its payload. However, for 
read requests, the situation is more involved, as the read requests 
bypass the AES core, and the inferred physical key context must be used 
to decrypt the read completions that will be returned by the host after 
the read request has been processed. 
%
%
When the PCI complex receives the read request, it caches the
\texttt{KEY\_INDEX} and \texttt{AES} fields in an on-chip lookup table along with other metadata, such as the source tile identifier. 
When the corresponding read completions arrive from the host, the PCI complex
retrieves these fields from this lookup tables
and inserts them into the read completion packets. 
The SXP can then use these values to identify the physical key context 
to use for decrypting the payload. 
The PCI complex also tracks the number of pending read completion packets for
each read request, and sets the \texttt{CC} bit on the last read completion packet. 

\section{Software Extensions}
\label{sec:sw}
We now describe a set of extensions to the IPU software stack---the XLA backend in Tensorflow, the Poplar compiler and the Poplar runtime---to
compile and execute confidential ML tasks using \sys\ in offline mode. 
This mode is triggered by a configuration option we have introduced in TensorFlow.
When this option is enabled, 
\begin{itemize}
\item The XLA backend transforms the computation graph to use a new abstraction called 
\emph{confidential data streams}(Section~\ref{sec:conf-streams}) for all data transfers; 
this includes initial weights, training data, checkpoints and the trained model. 

\item The Poplar compiler usually compiles the computation graph into
  a set of IPU application binaries (one for each IPU), where each binary is a
  concatenation of tile-specific binaries.
  In confidential mode, the compiler 
  encrypts each tile binary into a set of encrypted frames using a freshly
  sampled model key. Each frame is assigned a unique IV by composing the code type, 
  IPU ID, tile ID and frame index.
\item The Poplar runtime is extended to securely bootstrap the task,
  then transfer its encrypted application binaries and data between
  the host and the IPU (Appendix~\ref{appendix:sample} illustrates
  this process for a sample training scenario.)
\end{itemize}


\subsection{Confidential Data Streams}
\label{sec:conf-streams}

Confidential data streams are our compiler and runtime abstraction
for transferring data to/from the IPU with confidentiality and integrity guarantees,
leveraging SXPs.
Each stream is a sequence of data instances encrypted with the
same symmetric encryption key. 
Each data instance is partitioned into a sequence of frames, and each frame is
encrypted using a unique IV composed of a stream type (data), a stream
identifier (one for each stream in the application), and the index of the frame within the stream. 
%

IVs do not depend on application-specific attributes, such as 
batch sizes, positions in the IPU address space associated with the stream,
or the tiles that will issue read or write requests to the stream. 
Thus, a data stream can be encrypted and stored once, and then used for
training multiple models.

The compiler and the runtime implement reads and writes to
confidential data streams as follows. As discussed in
Section~\ref{sec:auth-enc}, the compiler first assigns a region in
tile PCI space to each stream, subject to the constraint that it never
exceeds the total capacity of the IPU ring buffer (e.g., 256 MB).

Next, the compiler assigns sets of tiles to read from or write to each stream,
reserves SRAM on each tile to hold a part of the stream, and generates SXP mappings,
subject to the constraints that (a) the exchange block context
associated with these tiles map to physical key contexts 
assigned to the stream, and (b) the number of physical key contexts in use at
any point in the program does not exceed 16 for any SXP. 
To maximize performance under these constraints, the compiler may
introduce synchronization points in the application where existing
keys are invalidated and new keys are loaded.
The compiler includes these synchronization points in the manifest,
along with their key identifiers; 
and the (untrusted) Popar runtime uses this part of the manifest 
to ask the CCU to load the next decryption keys into the SXPs at these points.

In multi-party scenarios, where tiles process encrypted data from 
multiple parties, the compiler may assign different sets of tiles to each
party's stream and introduce additional internal exchanges to distribute
the data to tiles that process the data altogether. 
This is due to the constraint that a tile is assigned only one physical key
context at a time, and thus cannot interleave accesses to data
encrypted with different keys.

The key changes apply only to input streams.
Keys for output streams are derived and loaded by the CCU
at the start of the TEE, and do not change throughput its lifetime.
Therefore, a malicious runtime that does not follow the key schedule
of the manifest can only cause decryption failures, resulting in denial-of-service.

Next, the compiler schedules read/write operations on each tile. 
The schedule is required to satisfy a hardware constraint that, at any point,
the tiles that generate requests targeting any given physical
key context be associated with a single exchange block context.
This is because, while the exchange block can
dynamically synchronize and regulate requests within each exchange
block context (so that its physical key context is used by one tile
at a time) there is no such synchronization across exchange block
contexts.

Finally, the compiler generates code on each tile that implements the schedule,
to issue read/write requests for the accessed frames.

For reads, the tile code (i) determines the size and expected IV of the next
frame; (ii) issues read request and wait for data to arrive in local
memory; (iii) checks that the IV contained in the frame matches the expected IV,
and generates security exception if not; (iv) strips IV and authentication tag;
(v) strips any application-level padding.
This sequence of steps is repeated for a pre-determined number of frames before
the communication phase ends and the computation phase begins.

For writes, the tile code repeats the following steps for as many frames as
needed to write the data: (i) determines the size and IV of the next frame;
(ii) writes the IV to the first 16 bytes of the current frame; (iii) splits the
frame into multiple packets, sets the \texttt{AES} bit in the header of all packets and the \texttt{CC} bit in the header of the last
packet, and adds padding to the last packet according to padding requirements; 
and (iv) issues the write requests for the frame.

\subsection{Secure Checkpointing}
Each IPU periodically checkpoints its state to enable recovery from failures. A checkpoint is created by writing the weights of the model to an output stream.
The checkpoint also includes metadata, such as the current offset for all
confidential data streams. These offsets are also written in plaintext, so that
the Poplar runtime can restart the job and resume loading of confidential data
streams at the correct offset.
Conversely, a checkpoint is restored by reading the weights using an input stream and resuming confidential streams from the checkpointed offsets. A checkpoint along with the job manifest and binaries suffice to resume an application from the checkpoint instead of restarting from the beginning.

In trusted mode, checkpoints are encrypted and integrity protected. In particular, tiles enforce the integrity of the process of restoring state from a previously created checkpoint. This includes protecting against attacks, such as tampering a checkpoint or loading a wrong checkpoint onto an IPU.
(Guaranteeing freshness, e.g., resumption
  from the latest checkpoint, would involve some form of trusted
  persistent storage and is out of scope in this paper.)

Checkpoints are implemented using confidential streams. 
The IV for each frame uniquely encodes the checkpoint type, the epoch
counter (incremented at each resumption), the checkpoint identifier
(incremented at each saved checkpoint), the IPU and tile IDs, and the
frame index.
The CCU uses a separate key for each epoch; it installs the key of the
epoch of the checkpoint it is resuming from, if any, and the key of
the current epoch for writing all its checkpoints.


The tiles read and write checkpoints as follows:
\begin{enumerate}[(1)]
\item Tiles obtain initial values of the epoch counter
  and checkpoint identifier from pre-determined locations in their tile
  memory; the CCU assigns them along with the bootloader code.
\item If the epoch counter is not null, tiles use it to
  compute their expected IVs and read their part of the corresponding
  checkpoint.

\item Each tile increments their local epoch counter and start (or
  resume) the application.

\item At regular intervals, the tiles checkpoint their part of the
  state, using IVs computed from their current values, and then
  increment their local checkpoint identifier.
\end{enumerate}

\subsection{Secure Bootstrapping}
\label{sec:bootstrapping}
Secure bootstrapping is the process of securely loading encrypted
application binaries into the IPU, either at the start of a job, or while
resuming a job from a checkpoint.

Bootstrapping involves the following steps. 
First, the Poplar runtime loads the encrypted IPU binary in host memory and
creates a TEE using the CCU APIs; this switches the IPU into trusted mode. 
Next, the CCU installs a \emph{bootloader} (shown in
Appendix~\ref{appendix:compiler}) onto every IPU tile using the autoloader described
in Section~\ref{sec:icu}, and also configures the SXPs with the model key and key regions in the tile PCI address space where the binary is loaded.
The bootloader on each tile fetches the tile's binary from host memory
by issuing a sequence of read requests. 
Each frame received from the host is intercepted by the SXPs, authenticated and
decrypted, and copied into tile memory. 
The bootloader then checks that the received IV matches the expected
IV built into the bootloader logic;
this check is performed in software because the SXPs only guarantee authenticity
of each frame, not the integrity of the entire code or data stream. 
The failure of this check indicates an attempt by the host to tamper with the code stream (e.g., by replaying or reordering frames). 
In such an event, the tile raises a security exception, which is handled by the
CCU. 
If all checks pass, the bootloader finally reconstruct the original cleartext binary by
stripping the IVs and authentication tags from all frames.

Finally, the bootloader computes a hash of the tile binary;
the tiles accumulate a hash of the whole application binary;
and the CCU checks that it matches the measurement in the
job manifest and generates a security exception otherwise.
This protocol, together with the integrity of the bootloader whose measurement
is included in the attestation report, guarantees application integrity. 

\subsection{Discussion: Online Mode and Inferencing}
\label{sec:sw-disc}
The software extensions described above can also be used in a configuration
where IPUs are coupled with a CPU TEE such as an Intel SGX enclave or AMD
SEV-SNP protected VM.
For example, an Intel SGX enclave can host TensorFlow, the Poplar compiler and
runtime, with an untrusted IPU driver running in the guest OS. 
In this configuration, the enclave would receive an encrypted model script from
the model developer, and the Poplar runtime would encrypt the compiled IPU
binary with fresh keys.
Similarly, the enclave would receive encrypted training data from data providers
on the basis of an Intel SGX and IPU attestation. 
The data can be decrypted, pre-processed and aggregated within the CPU enclave (in
parallel with job execution) and re-encrypted 
by the Poplar runtime with fresh keys. 
Encrypted code/data still then need to be copied to a run buffer allocated
outside the enclave (in the host process) accessible to the IPU. 
Inferencing can be supported in a similar way. 
We leave support for such scenarios as future work. 

\section{Evaluation}

Our evaluation focuses on the overhead of TEEs for ML training, using either CPUs or IPUs. 
TEEs for GPUs are discussed in the next section;
it would be hard to perform a precise comparison with their prototypes---we report on a first complete hardware implementation,  
whereas their experiments relied on simulations with benchmarks that are now outdated---but they exhibit 15--30\% overheads due to software encryption/decryption on GPUs, 
which our design eliminates using hardware encryption.
A general performance comparison with GPUs is out scope; see \cite{graphcore:perf}. 

\subsubsubsection{Implementation}
We have implemented ITX on a Graphcore GC200 IPU on a non-production development 
board. The IPU chip has been fabricated in TSMC's 7nm technology node, including 
the on-chip security extensions, which account for less than 1\% of the chip 
size. The CCU has been integrated on the development board and implements the 
firmware architecture described in Section~\ref{sec:trusted}, including the 
protocols for measured boot and TEE management.
As part of post-fabrication validation, the on-chip IPU security extensions have 
been tested to verify they conform to their specified behavior. We leave 
in-depth hardware security analysis and procedures that seek to defeat an
IPU TEE for future work.

We have implemented a software prototype for confidential training tasks where the host CPU server is untrusted, as discussed in Sections \ref{sec:overview} and \ref{sec:sw}. 
Our current prototype includes experimental support in the ML framework, Poplar compiler and runtime.
There are a few gaps in our prototype.
(1) Our implementation currently supports only one IPU on the board; 
(2) For simplicity, the compiler makes use of only one logical key region onto which code, data, label, checkpoints, and outputs are mapped; 
Nevertheless, every encrypted frame is statically assigned a unique IV, preserving the invariant that each IV is used only once; 
(3) Secure resumption is not yet implemented;
(4) The bootloader deployed on IPU tiles does not measure the application binary after authenticated decryption.

\subsubsubsection{Experimental Results} 
Table~\ref{tab:cifar10} summarizes the hardware and software configuration of
our test beds. We evaluate the performance of confidential training on ResNet
models of various sizes (20, 56, and 110) on the Cifar-10 dataset. The dataset
consists of 60,000 32x32 images spanning 10 classes; 50,000 of these images are
used for training the dataset and the remaining are used for testing the
resulting model.
We ran the same training code and data configurations in clear and confidential
mode, and confirmed that they both yield models with the same prediction
accuracy. 

We compare IPU TEEs against CPU TEEs based on the largest AMD SEV-SNP server we could find.
Since these are early development boards, we operate them at a reduced frequency
of 900 MHz; we expect better performance at higher frequencies.
The AMD CPU testbed utilizes 48 single-threaded AMD CPU cores (out of 64); 
Hyperthreading does not improve performance due to high vector unit utilization leaving little room for another hyperthread.
Moving from 32 to 48 cores improved performance by 10\%.

\begin{table}[t!]
    \begin{center}
        \begin{tabular}{p{0.17\textwidth}|p{0.27\textwidth}}
        \hline        \textbf{Testbed} &    
        \textbf{Training configuration} \\
        \hline
       	AMD SEV-SNP\newline
       	48-core VM on 64-core EPYC 7763 @ 2.4GHz & 
        ResNet-20. Batch size: 1534; 32 epochs.\newline 
        ResNet-56. Batch size: 768; 32 epochs.\newline
        ResNet-110. Batch size: 384; 64 epochs.\\
        \hline        
        Graphcore GC200 ITX\newline
        IPU @ 900 MHz, 2x24-core Intel Xeon 8168 & 
        ResNet-20. Batch size: 64; 32 epochs.\newline 
        ResNet-56. Batch size: 32; 32 epochs.\newline
        ResNet-110. Batch size: 16; 64 epochs.\\        
        \hline         
        \end{tabular}
    \end{center} 
    \caption{\small Testbed configuration for TensorFlow training of ResNet models on Cifar-10 dataset.
      In each configuration, batch sizes are optimized to to yield maximum performance. (Smaller batches do not affect correctness, but may improve convergence or accuracy.)}
    \label{tab:cifar10}
    \vspace{-0.2in}
  \end{table}

\begin{figure}[t]
    \includegraphics[width=3.4in]{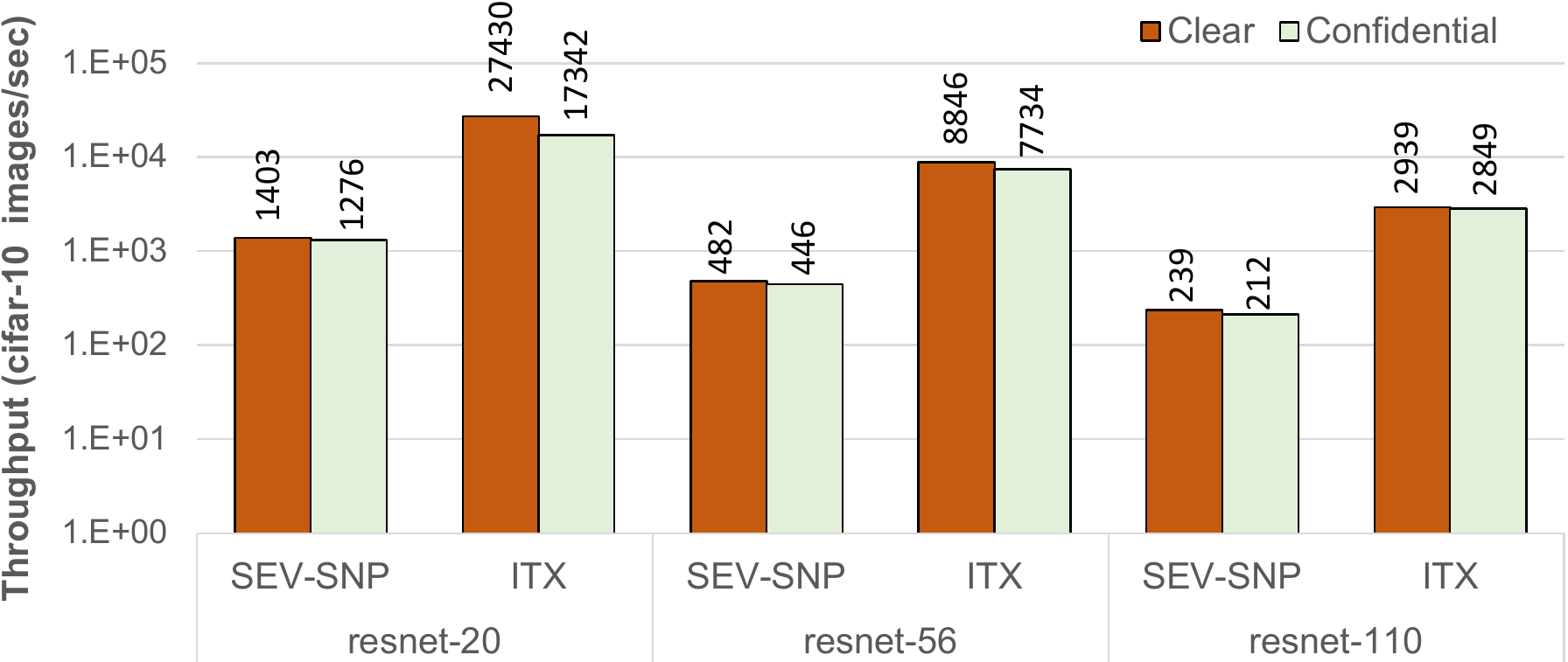}
    \center    
    \caption{\small Training throughput of ResNet models on cifar-10.}
    \vspace{-0.15in}
    \label{fig:performance}        
\end{figure}

Figure~\ref{fig:performance} shows the training throughput that we achieve in
clear and confidential modes. 
IPU-based training even with a single IPU operating at reduced frequency is
12-20x and 13-17x faster than CPU-based training in clear and confidential modes
respectively. 
Enabling SEV-SNP introduces modest overheads, ranging from 8\% (small model) to
14\% (large model) while the overheads of enabling \sys\ range from 3\% (large
model) to 58\% (small model). 
Virtually all the overheads of \sys\ can be attributed to the time spent in of
programming SXPs; the cost of encrypted I/O accounts for only 1.3\% of the overheads.
This is a temporary artifact of our prototype firmware, and can be
easily optimized. 
More generally, we expect the startup cost (TEE initialization, remote attestation, SXP setup) 
to be negligible with state-of-the-art models, which take weeks or days to train. 
With ResNet-110 model, the overall overhead is just 3\% (1123 vs 1089 seconds for running 64 epochs). 
We also expect that utilizing both IPUs at full frequency would deliver an
additional performance improvement (of up to 3.5x) over CPUs. 
In summary, the preliminary evaluation shows that using \sys, AI workloads can
continue to benefit from the use of accelerators without compromising on
performance or security. 

\section{Related Work}
\label{sec:related}

\subsubsubsection{ML Privacy} Machine learning involves many security
and privacy issues, which often need to be addressed both in their
application algorithms (applying, e.g., differential privacy and
federated learning) and in their system implementations.

Several interesting lines of work develop novel cryptographic schemes
for inference and training, relying on homomorphic encryption or
secure multi-party computation instead of trusted hardware. These
approaches can be implemented in software on existing CPUs,
and even benefit form GPU acceleration---see e.g. \cite{tramer:slalom,cryptgpu}.
They offer strong confidentiality, notably against side-channels.
However, they remain order-of-magnitude slower and more
resource-intensive than TEE-based approaches---see
\cite{hunt:chiron} for a comparison. They also require
significant algorithmic changes (to reduce the cost of fully
implementing floating-point operations and non-linear layers) and
separate mechanisms to protect the integrity of their computation.

\subsubsubsection{Trusted hardware}
There is a history of work~\cite{Evtyushkin2014IsoXAF, Boivie2011SecureBlueCS, Costan2016SanctumMH, Hofmann2013InkTagSA, McCune2009EfficientTR, Owusu2013OASISOA, TaMin2006SplittingIM, Lie2000ArchitecturalSF, Lie2000ArchitecturalSF,Weiser2017SGXIOGT} on trusted hardware that isolates code/
data from the rest of the system. Intel SGX~\cite{mckeen2013innovative} and AMD 
SEV-SNP~\cite{amd:sev-snp} are the latest in this line of work.
Our work effectively extends this approach 
from general-purpose CPUs to custom devices and accelerators. 

\subsubsubsection{Trusted execution on accelerators} 
To the best of our knowledge, our work is the first to demonstrate an ASIC
with confidential computing capabilities. NVIDIA recently announced confidential computing support in upcoming Hopper 
GPUs~\cite{nvidia:cc}. 
NVIDIA's design shares the same core principles as \sys\ on IPUs. 
The GPUs are equipped with an on-package hardware root of trust responsible 
for attestation and enforcing course-grained GPU isolation under the 
assumption that on-package GPU memory is trusted. 

There are some notable differences. For example, in the NVIDIA design, a 
CPU-based TEE capable of hosting a full OS is necessary because the 
responsibility of attesting and establishing a secure channel with 
the GPU lies with the GPU kernel-mode driver. Also, the performance 
characteristics of NVIDIA's approach are not yet known. Nevertheless, 
we believe that support for confidential computing in multiple accelerators
will greatly benefit the ecosystem.

Prior work has attempted to reduce trust on privileged host 
~\cite{Yu2015TrustedDO} via hardware support on the GPU \cite{svolos:graviton} or on the CPU \cite{insu:hix}. Graviton~\cite{svolos:graviton} extends the GPU with  support for secure resource management, and relies on a trusted GPU runtime  hosted in a process-based CPU TEE to manage the TEE lifecycle. HIX~\cite {insu:hix} provides extensions to process-based CPU TEEs, including the PCI interconnect and the CPU's MMU, to prevent system software from changing the PCI interconnect configuration and accessing GPU resources.

A number of researchers  have identified the need for mechanisms that allow an application hosted in a TEE to securely communicate with I/O devices, such as 
in-storage processors ~\cite{kang:iceclave}, GPUs~\cite{insu:hix, 
svolos:graviton, Yu2015TrustedDO}, FPGAs~\cite{khawaja:amrophos, oh:meetgo, 
pereira:towards, zhao:shef}, and AI accelerators~\cite{hua:guardnn, xie:ai}. 

GuardNN removes the CPU from the TCB of AI systems~\cite{hua:guardnn}. 
It introduces instructions for establishing a 
secure channel between a remote user and the accelerator, and for decrypting/ 
encrypting inputs/outputs. Integrity is not guaranteed as an 
attacker controlling the CPU can tamper with the instruction schedule. 

HETEE ~\cite{zhu:hetee} enables confidential rack-scale AI computing using a 
tamper-resistant chassis that consists of computing nodes, commodity 
accelerators, a PCI switch, and a security controller. The security 
controller enables remote attestation and remote users to establish a 
secure channel with the chassis.

\subsubsubsection{PCIe-level encryption}
The deployment of devices that support standardized PCIe-level encryption \cite{pcisig:ide}
is expected to start in a few years.
Compared to application-level encryption (Section~\ref{sec:integrity}) 
it may enable a more transparent and more efficient CPU--IPU protocol (removing the need for explicit IVs in ML applications.)
However, it would involve a larger TCB with an auxiliary host CPU TEE.


\section{Conclusion}
We presented ITX, a set of experimental hardware extensions for the Graphcore IPU, a state-of-the-art AI accelerator.
Our design provides application-level confidentiality and integrity for ML tasks offloaded to an untrusted cloud provider.
We also presented a software architecture that avoids the need to trust host CPUs,
thereby minimizing the trusted computing base and removing dependencies on CPU-based TEEs.
We implemented them in the Graphcore GC200 IPU,
and experimentally confirmed small performance overheads for training large models with strong security and privacy.

\vfill

\bibliographystyle{ACM-Reference-Format}
\bibliography{references}

\appendix
\section{APPENDIX}
\subsection{Attack Vectors and Security Analysis}
\label{app:attack-analysis}

Table~\ref{tab:attack-analysis} summarizes the attack vectors discussed in Section~\ref{sec:threat}
and, for those covered by our threat model, how ITX mitigates each of these attacks. 

\begin{table}[th]
  \small
  \begin{center}
    \begin{tabular}{p{0.21\textwidth}|p{0.23\textwidth}}
      \hline
      \textbf{Threat} & \textbf{Mitigation}  \\ \hline
      
      \textbf{Host (Software, Physical)}  & {} \\ 
      {\textit{IPU Memory Access}, 
       e.g., host software uses MMIO
       and PCI BARs, physical attacker
       tampers with on-chip memory}       & {MMIO blacklist prevents CPU from 
                                             accessing code and data in IPU;
                                             access via interfaces like JTAG is prohibited;
                                             IPU memory cannot be physically accessed 
                                             without breaking into the package.} \\[1ex]
       {\textit{Host CPU, Memory, and PCIe bus}, e.g., 
       read, write, replay, or
       re-ordering of code and data in 
       host memory or in transit,
       including DMA buffers and PCIe bus}   & {Code and data are encrypted with AES-GCM 
                                             using explicit IVs, and keys not shared with the host;  
                                             uniqueness and integrity of IVs are ensured 
                                             by trusted code executed on tiles.} \\[1ex]                                                 
      {\textit{IPU Binary Malleability},  
       e.g., host replaces model 
       encryption key or encrypted
       code}                             & {Bootloader computes hash of the tile binary;
                                            hash accumulated and checked against expected 
                                            measurement in the job manifest. (Not implemented 
                                            in the evaluated prototype.)}\\[1ex]                                                 
      {\textit{IPU Connectivity}, } 
      ICU-CCU or ICU-IPU Tampering on the development board & {\textbf{no}; attacker can mount a physical attack to
                                             (1) retrieve the key(s) sent to IPU, and 
                                             (2) tamper with ICU firmware measurement sent to CCU} \\[1ex]
      IPU-IPU Tampering                   & {\textbf{no}; attacker can mount physical attacks against multi-IPU tasks by tampering with data sent 
                                             between IPUs.}\\                                        
      \hline
      \textbf{Supply Chain and Firmware}               & {} \\      
      {Primary Bootloader Provisioning
       Tampering}                         & {Graphcore checks whether the signed bootloader 
                                             manifest includes the expected nonce provisioned
                                             into the CCU primary bootloader.} \\[1ex]
      {Using non-genuine, known vulnerable TCB components} & {Firmware authorization; hardened measurement 
                                             protocol outlined in 
                                             Appendix~\ref{sec:hardened-protocol}.} \\      
      \hline

      \textbf{Side-channels}          & {} \\
      {IPU Memory}                    & {IPU memory access patterns cannot be observed by
                                         co-located attacker as the IPU is entirely assigned to 
                                         one job at a time.} \\[1ex]
      {Host Memory and PCIe bus}          & {\textbf{no}: attacker can observe access patterns
                                         to host memory and on PCIe bus. However, these patterns do not leak much information in the BSP model, e.g., the size and number of minibatches, but not their contents.} \\[1ex]
      {Power- and timing-level}       & {\textbf{no}: attacker can measure power consumption
                                         and/or execution time of a superstep. Similarly, this does not leak much information for typical ML tasks.} \\
      \hline
    \end{tabular}
    \end{center} 
    \caption{\small Potential threats and how ITX mitigates them. Physical access 
    attacks on the CCU-ICU-IPU and the IPU-IPU channels can be mitigitated 
    once the CCU is integrated on the IPU and AES-GCM is utilized to protect the IPU-IPU 
    channels.}
    \label{tab:attack-analysis}
\end{table}

\subsection{Firmware Provisioning and Device Certification}
\label{app:fw-prov-certification}
In this section, we describe an example process for firmware provisioning and device 
certificates that would be followed if \sys{} were to be used in a production Graphcore IPU 
products in a production environment.
During board manufacturing, the CCU would be provisioned with firmware followed by a board 
reset to harvest certificate signing requests (CSRs) generated by the execution of the 
primary and secondary bootloaders. The CSRs would then be used by Graphcore to issue device certificates.

\subsubsubsection{Firmware Provisioning}
The CCU is provisioned with firmware using the SoC's Secure Firmware Install (SFI) feature~\cite{stmicro:sfi}. The firmware package consists of all firmware layers discussed in Section~\ref{sec:rot} and the configuration bytes (called \texttt{OPTION}), whose secure user memory registers are configured so that secure user memory includes only the regions onto which the secure bootloader is deployed. 
The firmware package is encrypted with a symmetric key, which is provisioned to a hardware security module (HSM). 

The encrypted firmware package and the HSM are used by the board manufacturer to deploy CCU firmware during the manufacturing and testing of the Graphcore products. The chip tester implements a multi-stage protocol between the CCU secure bootloader and the HSM, during which the HSM authenticates the certificate issued by the CCU and wraps its firmware encryption key using the certified public key. This enables the CCU secure bootloader to decrypt the firmware package, to install the firmware, and to configure the \texttt{OPTION} bytes based on the requested configuration.

While this SFI process guarantees confidentiality of the firmware, it does not directly protect its integrity: the provisioning process may be subject to supply-chain attacks that would replace CCU parts provisioned using SFI with CCU parts containing malicious firmware. We extend SFI with protection against such attacks by injecting a secret known only to Graphcore into the primary bootloader. Once the CCU has been integrated onto a Graphcore board, a challenger can ask the primary bootloader to prove possession of the secret. 

This process entails the following three steps. First, Graphcore generates a fresh secret for every batch of CCUs. The secret is injected to the primary bootloader of the CCU firmware. Second, Graphcore derives from the secret an asymmetric batch-specific bootloader manifest signing key. After deriving this key, Graphcore keeps only the public part. Third, Graphcore issues a certificate for the public bootloader manifest signing key. The certificate is signed by the Graphcore Firmware certificate authority (CA). This certificate contains a batch number, and is valid till the production date of the batch of CCUs.

\subsubsubsection{Device Certification}
In order to certify its device identity keys, 
the board tester resets the board and harvests the CIK and PIK CSRs generated for the board and platform identity keys, as well as the bootloader manifest. The command to harvest the bootloader manifest includes a fresh nonce, to be echoed in the signed bootloader manifest.

In response, Graphcore verifies the CSRs received by the card manufacturer and issues CIK and PIK certificates that are signed by the Graphcore CIK and PIK CAs. In addition, Graphcore validates the bootloader manifest against the bootloader manifest signing key certificate expected specific to the batch to which the CCU belongs, and ensures that the nonce matches the expected nonce. 

\subsection{Firmware Updates}
\label{app:fw-sign-updates}
The CCU firmware consists of mutable secondary bootloader and CCE,
both authenticated by the primary bootloader and possibly updated
after the Graphcore card has been deployed in production.


\subsubsubsection{Updates to Secondary Bootloader}
The secondary bootloader involves relatively complex cryptographic opertions, and may need to be updated in the field.
As discussed in Section~\ref{sec:attestation},
the platform identity key (PIK) is derived from UDS depending on the hash of the secondary bootloader. Therefore, any updates to the secondary bootloader changes the platform identity, and PIK certificates issued by the manufacturer are no longer valid, requiring re-certification of the device by the manufacturer. 

Unfortunately, re-certification of a remote device by the manufacturer can be a complex and lengthy operation as the manufacturer (by design) does not retain unique device keys. Thus, it requires collection of CSRs from the device, and more importantly an authentication mechanism to ensure that the manufacturer signs only PIK certificates exported from devices in the cloud provider's datacenters. 

We overcome this challenge via a protocol that enables updates to the secondary bootloader without invalidating manufacturer-issued certificates. 

Prior to updating the secondary bootloader (say to version Y), the cloud provider's Graphcore Firmware CA issues a TCB update certificate capturing the measurement of the new version of the secondary bootloader and revokes previous certificates for versions of the secondary bootloader that should no longer be deployed.

After a firmware update has been deployed, the primary bootloader generates a new CDI ($CDI^Y$). The secondary bootloader generates platform identity and attestation keys specific to this version of firmware ($PIK^Y$ and $AK^Y$). However, the card identity key (CIK) stays the same as it does not depend on the measurement of the secondary bootloader. The $PIK^Y$ certificate, hence, is signed by the original CIK, which has been certified by the manufacturer.

Subsequently, a remote challenger can combine the TCB update certificate with the CIK certificate originally issued by the manufacturer to verify the $PIK^Y$ certificate is issued by the device using the original CIK, and that the measurement of the new secondary bootloader in the $PIK^Y$ certificate matches the measurement of the secondary bootloader in the TCB update certificate.

\subsubsubsection{Updates to CCE} They can be applied at any point without the need for any additional certification from the manufacturer. When a device boots with a new version of CCE, it generates a new attestation key with a signature over the public AK along with a hash of the CCE using the PIK. Quotes generated by the updated version of CCE firmware can be validated using a valid PIK certificate.

\subsection{Measured Boot Protocol}
\label{sec:hardened-protocol}
\begin{figure}[t]
  \centering
  \includegraphics[height=4in]{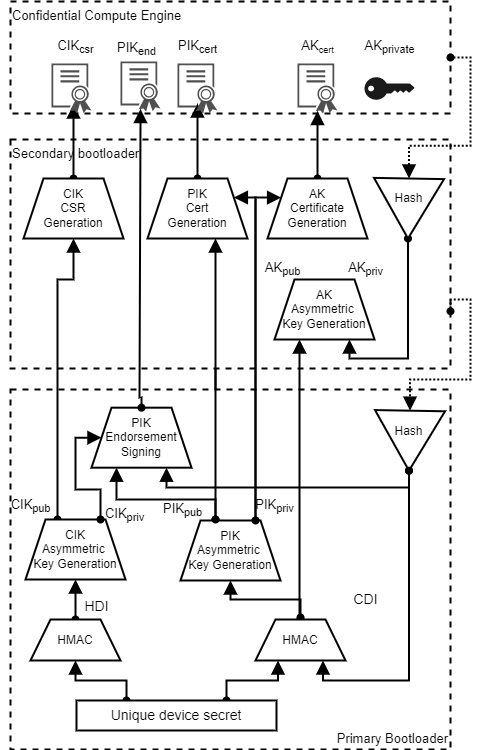}
  \caption{\label{fig:firmware_arch_alt}\small Hardened boot protocol that protects against bootloader impersonation attacks.}
\end{figure}

As discussed in Section~\ref{sec:rot}, the measured boot protocol is susceptible to advanced impersonation attacks. 
We can harden the boot protocol further by moving CIK and PIK generation into
the primary bootloader (as shown in Figure~\ref{fig:firmware_arch_alt}) without
revealing the private CIK to the secondary bootloader. 

In this protocol, the primary bootloader generates CIK from UDS, and generates
PIK using CDI and the measurement of the secondary bootloader. 
To allow a relying party (such as the Graphcore CA) to attest that the PIK was
indeed generated by the primary bootloader, the primary bootloader creates a
custom structure known as PIK endorsement containing the PIK public key along
with a measurement of the secondary bootloader, and a signature over these two
attributes using the CIK.
The bootloader then scrubs the CIK private key and passes public CIK and PIK
keys along with the private PIK and the PIK endorsement to secondary bootloader.
During manufacturing, the Graphcore PKI issues a PIK certificate only after
validating the PIK endorsement structure. 

Our prototype CCU does not implement this protocol to keep the primary
bootloader simple, but this is easy to address in the future. 

\subsection{Compiled Manifests and Bootloader}
\label{appendix:compiler}
\subsubsubsection{Job Manifest} The Poplar compiler generates a job
manifest, which includes all the information required by the Poplar runtime and CCU to 
create and launch a new TEE, which will host the ML task. The manifest contains 
the hash digest of the application binary loaded into each IPU. 
Finally, it lists the synchronization barriers (or points) at which the IPU needs to synchronize with the host, and for each synchronization point, it keeps the following information:
\begin{itemize}
	\item the key region identifier assigned to each stream that will be read or written 
	following the end of synchronization (i.e., the mapping between a stream identifier $j$ to a key region identifier);
	\item the ring buffer region (i.e., Tile PCI space in the ring buffer) 
	assigned to each key region (key region definition registers);
	\item the part of each stream that has been mapped to the ring buffer region 
	(stream offset);
	\item the set of physical key contexts to which the stream key needs to be loaded;
	\item the physical key context assigned to each exchange block context (exchange block context map registers); and
	\item the key region to which each physical key context is assigned (physical key map registers).
\end{itemize}

\subsubsubsection{Secure Code Bootstrapping}
The code snippet below illustrates the bootloader code that fetches the frames of an application binary and confirms the integrity of the IV of each frame.

\begin{lstlisting}
def bootloader():
  ipu_id = get_current_ipu_id()
  tile_id = get_current_tile_id()
  num_frames = TOTAL_TILE_MEMORY / (MAX_FRAME_SIZE - IV_SIZE -TAG_SIZE)

  for index in range(1, num_frames):
    expected_iv = StreamType::CODE | ipu_id | tile_id | index
    frame = read_next_frame_from_host()
    if expected_iv != get_iv(frame):
      raise_security_exception()
    strip_iv_and_tag(frame)
  compute_hash()
\end{lstlisting}

\subsection{Attestation}
\label{appendix:job}

\subsubsubsection{Cryptographic Operations}
Table~\ref{tab:crypto} details the keys 
sampled, derived, and exchanged at the start of a run in trusted mode.
We rely on standard algorithms: Elliptic Curve Diffie-Hellman for establishing
shared secrets, a KDF for deriving keys, and an AES-based authenticated key-wrapping scheme.
These operations rely on the attestation of the manifest and runtime parameters (including all public keyshares).
Each party provides its own random nonce, and the CCU combines them to
deterministically derive keys for checkpoints and the final
model; these keys are fresh secrets as long as one party is honest.
In order to resume from a checkpoint saved in a previous run, the
attested runtime parameters ensures that all parties agree on the
epoch and checkpoint identifiers, and the parties provide their nonces
both for the previous run and for the new run.

\begin{table}[t]
  \small
  \begin{center}
    \begin{tabular}{@{}p{0.175\textwidth}@{}p{0.04\textwidth}|p{0.11\textwidth}|p{0.11\textwidth}@{}}
      \hline
      \textbf{Key or secret} & \textbf{} & \textbf{Provider} & \textbf{CCU} \\ \hline
      public/private keyshare \newline for each relying party $p$ \!\!\!\!            & $X_p, x_p$ &  fresh EC share  & receive $X_p$ \\  \hline
      
      encryption key \newline for each input stream $j$    & $k_j$      &  fresh key & unwrapped \\ \hline
      public/private keyshare \newline for the CCU in this run  & $Y, y$     &  receive $Y$  &  fresh EC share \\  \hline
      nonce for $p$ in this run        & $s_{p,Y}$  &  fresh secret & unwrapped \\  \hline
      wrapping key \newline  for $p,Y$ with salt \newline
      $a = X^p||Y||M$        & $w_p$      & ${\sf KDF}[x_p\cdot Y](a)$ & ${\sf KDF}[y \cdot X_p](a)$ \\ \hline
      key to load checkpoints \newline saved by prior run $Z$  & $k_{\textit{load}}$ & N/A & ${\sf KDF}[\vec{s}_{p,Z}](\textit{'ck'})$ \\ \hline 
      key to save checkpoints & $k_{\textit{save}}$ & N/A & ${\sf KDF}[\vec{s}_{p,Y}](\textit{'ck'})$ \\ \hline 
      key to save final model &     $k_{\textit{m}}$ & unwrapped & ${\sf KDF}[\vec{s}_{p,Y}](\textit{'m'})$
    \end{tabular}
    \end{center} 
    \caption{\small Keying for a workload with manifest $M$ between relying parties identified by their public keyshares $\vec X_p$
      and a CCU identified by its fresh CCU public keyshare $Y$ for this run. 
      After attestation,
      an ECDH shared secret $w_p$ is used for wrapping $k_j$, $s{_{p,Y}}$, and $s{_{p,Z}}$ when resuming from $Z$
      from $p$ to the CCU, and optionally for wrapping $k_{\textit{m}}$ from CCU to any party $p$ designated as a receiver of the final model.
      The keys used for encrypting checkpoints and the final model are derived from nonces from all relying parties, ensuring
      these keys are fresh (as soon as one party is honest) and require agreement from all parties to be released.
    }
    \label{tab:crypto}
\end{table}

\subsubsubsection{Remote Attestation}\label{sec:attestation-details}
During TEE creation, 
the CCU generates an attestation report that captures
security-critical attributes 
about the IPU and runtime configuration, including 
\begin{itemize}
\item the measurement of configuration registers;
\item the measurement of the IPU bootloader used for loading application binaries onto IPUs;
\item the measurement of the job manifest;
\item the hash digest of the attributes for this run, including:
  \begin{itemize}
  \item the public keyshare of the CCU for this run ($Y$);
  \item the epoch $e$ and checkpoint counter $c$ from which the job is restarted (if any).
  \item the certificate fingerprints of all parties ($\vec{X}_p$);
  \item a stream assignment, specifying a party for each input, 
  and parties (model receivers) that receive the model key;
  \end{itemize}
\end{itemize}

The host collects the attestation report, along 
with a CCU-issued certificate chain, which includes the AK, PIK and 
CIK certificates, and is rooted at the self-signed CIK certificate. 
These are presented to relying parties along with: 
the original CIK and PIK certificates, 
the TCB update certificates for the secondary bootloader and ICU firmware,
and any intermediate CA certificates. 

A relying party can verify the attestation report as follows:
\begin{enumerate}[(1)]
\item Validate the CCU-generated certificate chain and auxiliary
  certificates. This includes checking for certificate revocation.

\item Confirm that public key of the CIK certificate issued by
  Graphcore matches the public key in the CIK certificate obtained
  from the CCU.

    \item Confirm that any updates to the secondary bootloader
    and ICU firmware are rooted to a valid certificate chain,
    In doing so, two checks are required: (i) 
    if there exists a TCB update certificate issued for the secondary 
    bootloader with a hash digest matching the hash digest in the CCU-
    issued PIK certificate; 
    (ii) if there exists a TCB update certificate issued for ICU 
    firmware with a hash digest matching the hash digest in the CCU-
    issued PIK certificate.

  \item Review the attested manifest and attributes for this run.

\end{enumerate}

\subsubsubsection{Secure Key Exchange} \label{appendix:attestation-sk}
For each run, each party $p$ derives a fresh wrapping key $w_p$ using its private keyshare 
$x_p$ and the public keyshare of the attested CCU $Y$. This key is used to wrap a key package 
containing the streams identifiers assigned to the party and the party's key for these 
streams $k_j$, and the nonce(s) $s_{p,Y}$ for the current run (and $s_{p,Z}$ for the previous run if the current run is resuming from a checkpoint saved in run $Z$.) The CCU can derive the wrapping key for party 
$p$ using its private keyshare $y$ and the party's public keyshare $X_p$. In 
possession of $\vec w_p$, the attested CCU can unwrap the key packages 
of all parties, which are made available during the TEE launch stage. 

The parameters of the model are encrypted using the final-model key $k_m$ that 
has been derived by the CCU using the nonces obtained from all parties. The parties engage in a protocol for exchanging their nonces so they can derive the key once they possess all nonces. The CCU can additionally release the final-model key to model receivers listed in the attestation report using the wrapping key shared between itself and each model receiver.

\subsection{Sample Training Scenario}
\label{appendix:sample}
\begin{figure}[th]
  \includegraphics[width=3.4in]{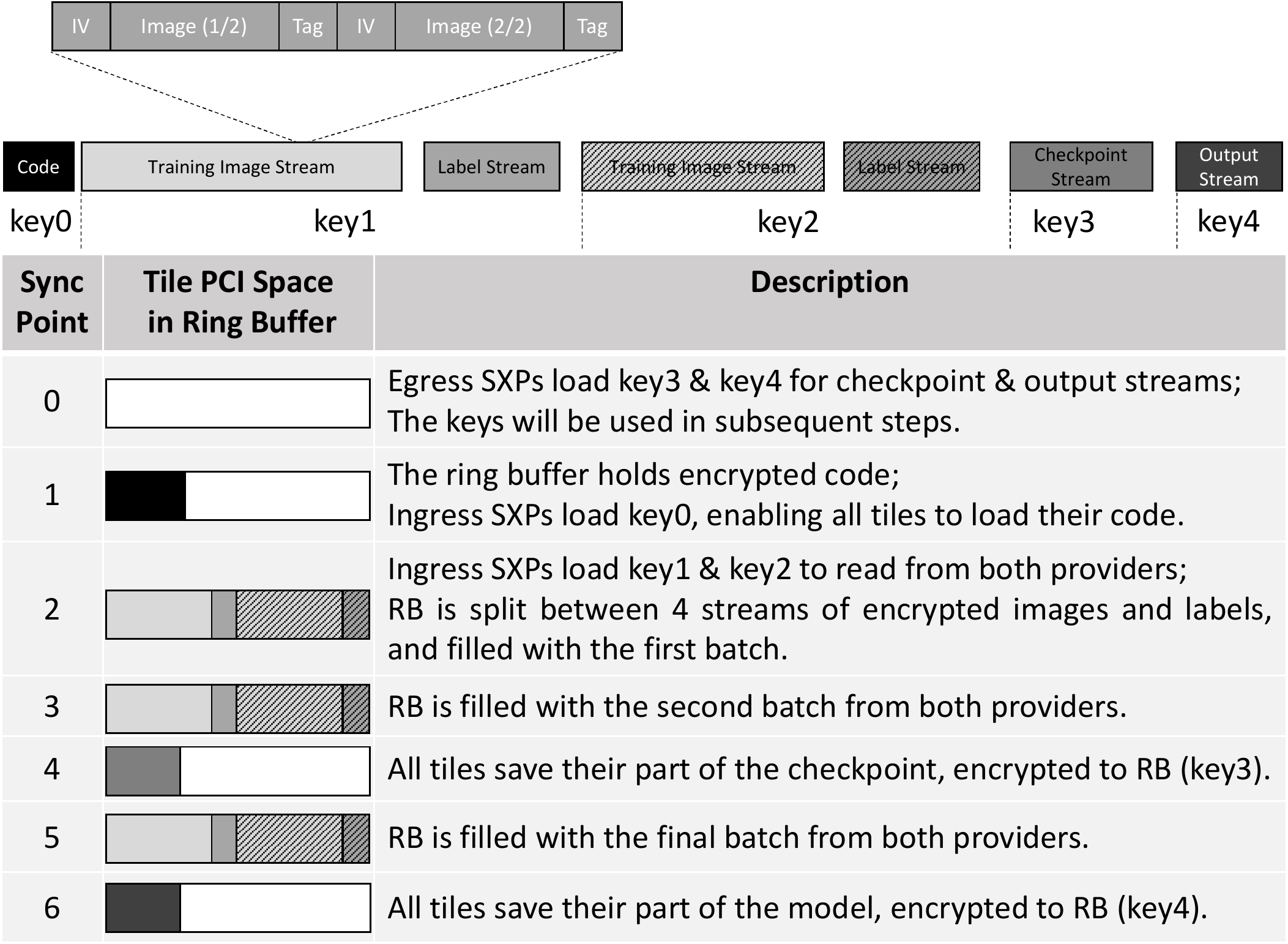}
  \center
  \vspace{-0.1in}
  \caption{\small Sample training scenario with 3 parties: one providing
    model code (using key0) and the others (using key1 and key2)
    each providing their own streams of training images and labels;
    this task saves checkpoints (using key3) and a final model
    (using key4).
    The compiler emits a job manifest that
    indicates, for each synchronization point of the task,
    which part of each stream is mapped to the ring buffer (1..6)
    and which keys the CCU should load for ingress. 
    The keys for egress streams are programmed in the start of the job (0).
  }
  \label{fig:conf-streams}        
  \vspace{-0.15in}
\end{figure}

Figure~\ref{fig:conf-streams} illustrates a sample training scenario with 
three parties. Given the job manifest generated by the Poplar compiler, Poplar 
runtime, CCU, and IPU synchronize at various points where the Poplar runtime 
populate the ring buffer with the data expected by the IPU, and the CCU loads
enryption keys to the IPU SXPs.

\end{document}